\documentclass[preprintnumbers,amsmath,amssymb]{revtex4}
\usepackage{amsfonts}
\usepackage{graphics}
\usepackage{epsfig}
\usepackage{url}
\usepackage{multirow}
\newcommand {\be}{\begin{equation}}
\newcommand {\ee}{\end{equation}}
\newcommand {\ba}{\begin{eqnarray}}
\newcommand {\ea}{\end{eqnarray}}
\newcommand {\invfb}{$fb^{-1}$}
\newcommand {\tanb}{tan$\beta$}
\newcommand {\ra}{\rightarrow}

\begin{document}
\title{Possibility of observing MSSM charged Higgs in association with a W boson at LHC}
\author{M. Hashemi}

\affiliation{School of Particles and Accelerators, Institute for Research in Fundamental Sciences (IPM),
P.O. Box 19395-5746, Tehran, Iran}

\begin{abstract}
Possibility of observing associated production of charged Higgs and W boson in the framework of MSSM at LHC is studied. Both leptonic and hadronic decays of W boson are studied while the charged Higgs boson is considered to decay to a $\tau$ lepton and a neutrino. Therefore two search categories are defined based on the leptonic and hadronic final states, i.e. $\ell~\tau+E^{miss}_{T}$ and $jj~\tau+E^{miss}_{T}$ where $\ell=e$ or $\mu$ and $j$ is a light jet from $W$ decay. The discovery chance of the two categories is evaluated at an integrated luminosity of 300 \invfb~at LHC. It is shown that both leptonic and hadronic final states have the chance of discovery at high \tanb. Finally $5\sigma$ and $3\sigma$ contours are provided for both search categories. 
 
\end{abstract}

\maketitle

\section{Introduction}
The Standard Model of particle physics has obtained a magnificent success through the experimental precision tests in the last years.
Despite its success, the main part of the theory, i.e. the Higgs mechanism and the question of mass spectrum has remained unexplored and out of reach by the recent experiments. One of the main goals of upcoming experiments especially the Large Hadron Collider experiment (LHC) at CERN is to search for the Higgs boson as the predicted particle through the Higgs mechanism. While prediction of a single Higgs boson in the SM appears to be incomplete and reveals theoretical problems such as the Higgs boson mass divergence when including radiative corrections, theoretical models beyond the SM appear to be attractive candidate solutions to the SM Higgs sector problems. The supersymmetric extensions to the SM are one of these models which not only provide elegant solutions to the Higgs boson mass divergence by introducing supersymmetric partners for SM particles but also provide approaches for other issues such as the gauge couplings unification Ref. \cite{Martin}. The family of two Higgs doublet models (2HDM) which introduce two Higgs doublets instead of a single scalar Higgs particle are of special interest among the supersymmetric models. The minimal supersymmetric extension to the SM, the so called MSSM, appears as a simple, reliable and minimal model belonging to the 2HDM family which predicts five physical Higgs bosons two of which are charged. While the neutral Higgs bosons of MSSM might be hard to distinguish from neutral Higgs boson of the SM, the observation of a charged Higgs is a crucial signature of theories beyond the SM. This is an enough reason that this particle has attracted special attention in the last years in different High Energy Physics experiments and will certainly be probed at the LHC experiment at CERN.
\\
The current results on the search for the charged Higgs include the low mass search by the LEP Higgs Working Group which excludes a charged Higgs with $m_{(H^{+})}<80$ GeV in a direct search in Ref. \cite{lepexclusion1} while the combined result of MSSM Higgs boson searches excludes a light charged Higgs with $m_{(H^{+})} < 125$ GeV in the $m_{h}-max$ scenario in Ref. \cite{lepexclusion2}. The high \tanb$~$ region has been excluded by the CDF collaboration quoted in Ref. \cite{cdfexclusion}.\\ 
At LHC a large number of searches have been carried out in CMS and ATLAS collaborations. These searches divide the parameter space into two regions of light and heavy charged Higgs. The light charged Higgs ($m_{(H^{+})}<175$ GeV) is produced from the top pair production while the heavy charged Higgs ($m_{(H^{+})}>175$ GeV) is produced through $gg\ra t\bar{b}H^{-}$ and $gb\ra tH^{-}$ with a combination procedure described in Refs. \cite{TPlehn} and \cite{TPlehn2}.\\
In Ref. \cite{LCHCMS} a light charged Higgs has been studied in CMS in the leptonic final state i.e. $H^{+}\ra \tau^{+}\nu$, $W\ra\ell\nu$, with $\ell=e$ or $\mu$. In Ref. \cite{HCHCMS}, a heavy charged Higgs has been analyzed in CMS in the hadronic final state, i.e. $H^{+}\ra \tau^{+}\nu$, $W\ra jj$. Both analyses predicted a possible discovery for a wide region of parameter space at an integrated luminosity of $30 fb^{-1}$. The $H^{+}\ra t\bar{b}$ decay channel has been analyzed in CMS in Ref. \cite{LowetteCH} however it was shown to be suffering from the large hadronic background.\\
The ATLAS collaboration results are reported in Ref. \cite{CHATLAS}. These analyses include three main final states for the light charged Higgs i.e. the hadronic decay of the $\tau$ lepton with leptonic or hadronic decay of the $W$ boson, and the leptonic decay of the $\tau$ lepton with hadronic decay of the $W$ boson. The heavy charged Higgs has also been analyzed by the ATLAS collaboration looking at $H^{+}\ra \tau^{+}\nu$ and $H^{+}\ra t\bar{b}$. These analyses provide discovery and exclusion contrours corresponding to integrated luminosities of 1, 10 and 30 $fb^{-1}$ on a significant fraction of the parameter space. There are also early data searches such as Ref. \cite{CH_ED_ATLAS} which predict possible extensions to the current D0 results in Tevatron with data as early as a time needed to collect $200pb^{-1}$ integrated luminosity with $\sqrt{s}=10$ TeV collisions. Although both $H^{+}\ra \tau^{+}\nu$ and $H^{+}\ra t\bar{b}$ decays are considered for heavy charged Higgs search, the latter does not lead to a promising discovery chance leaving the attention to be paid to the tauonic decay of the charged Higgs as the main decay channel throughout the parameter space. Therefore in the analysis presented in this paper only $H^{+}\ra \tau^{+}\nu$ is analyzed. 
\\
The associated production of charged Higgs and a W boson is a complementary search channel to the current charged Higgs analyses. This channel is attractive in the transition region where the on-shell $t\bar{t}$ production switches to the $gg\ra t\bar{b}H^{-}$ and $gb\ra tH^{-}$ processes with the possibility of off-shell production of top quarks. Since the signal cross section decreases in the transition region around the top quark mass, the experimental selection of signal events is a challenge in this region. However since the $H^{+}W^{-}$ process can occure in both light and heavy charged Higgs area, it can help increasing the signal sensitivity especially in the transition region. \\
This channel has been extensively studied in the literature from theoretical and phenomenological points of view. In Refs. \cite{24} and \cite{25} the main production diagram was shown to be the tree level $s$-channel and $t$-channel $b\bar{b}\ra H^{\pm}W^{\mp}$ while the sub-dominant production process i.e. the one-loop diagram $gg\ra H^{\pm}W^{\mp}$ which proceeds through the heavy quark triangular and box loops has a negligible contribution especially in the low charged Higgs mass region and high \tanb$~$ values. \\
In Refs. \cite{27} and \cite{28} quark and squark loops contributions to the $H^{\pm}W^{\mp}$ production process were evaluated and shown to contribute a little to the $gg\ra H^{\pm}W^{\mp}$ process keeping the $b\bar{b}\ra H^{\pm}W^{\mp}$ still dominant. In Ref. \cite{29} different choices of soft supersymmetry breaking parameters with or without squark mixing were studied and their effects on the $gg\ra H^{\pm}W^{\mp}$ process were evaluated. Again although these scenarios can have sizable effects on $gg\ra H^{\pm}W^{\mp}$, they were found to keep the $b\bar{b}\ra H^{\pm}W^{\mp}$ process as the dominant process. The supersymmetric electroweak corrections were also studied in Ref. \cite{30} and they were shown to be negligible in the high \tanb$~$ region where \tanb$\simeq30$. In Ref. \cite{31} the $O(\alpha_s)$ QCD contributions including virtual corrections as well as real gluon radiations were calculated in $\overline{MS}$ and $OS$ schemes and the correction to the tree level process $b\bar{b}\ra H^{\pm}W^{\mp}$ was estimated to be approximately $-15\%$ for the low mass charged Higgs and high \tanb. In Ref. \cite{32} supersymmetric QCD corrections were calculated and compared with corresponding SUSY-EW corrections (Ref. \cite{30}) and $O(\alpha_s)$ QCD corrections (Ref. \cite{31}). As a result as an example at \tanb=40 it was shown that SUSY-QCD corrections are positive and can decrease $O(\alpha_s)$ QCD and SUSY-EW corrections by $\sim50\%$ for $m_{(H^{+})}=200$ GeV. \\
The NLO QCD corrections to the cross section of this process were also calculated in Ref. \cite{33} and the results indicate that the K-factor (defined as $\sigma_{NLO}/\sigma_{LO}$ where $\sigma_{NLO}(\sigma_{LO})$ is the total cross section with NLO(LO) QCD corrections included) is close to unity for \tanb=40 and $m_{(H^{+})}=200$ GeV. Possible enhancement of the production cross section has also been studied in Ref. \cite{34} in a general 2HDM and results show that there are regions in the parameter space where the 2HDM predicted cross section could be larger than that predicted by MSSM by two orders of magnitude.\\
There have been several phenomenological analyses of this channel based on Monte Carlo simulation and event selection in the literature. In Ref. \cite{35} the charged Higgs decay to a top and a bottom quark through $H^{+}\ra t\bar{b}$ was analysed and it turned out that there is no chance of signal extraction from the large hadronic background with this decay channel. Alternatively in Refs. \cite{36,37,38} the $H^{+}\ra \tau^{+}\nu_{\tau}$ decay channel was studied in the hadronic final state i.e. with $W\ra jj$ and the $W+2$ jets process was taken as the main background in the analysis and concluded that in parts of the parameter space it is possible to have a signal significance exceeding $5\sigma$. \\
The current analysis aims at three main purposes complementary to the previous analyses: first including a set of main background processes in the analysis i.e. $t\bar{t}$, $WW$ and $W+$jets and estimating their contribution to the signal region, second extending the search to the leptonic final state which involves leptonic decay of W boson (i.e. $W\ra \ell \nu$ where $\ell = e$ or $\mu$) and third extending the search to the light charged Higgs area where $m_{H^{+}}<175$ GeV. The analysis starts with heavy charged Higgs which is studied in two categories of leptonic and hadronic final states. Then the light charged Higgs signal is analyzed and two production processes, i.e. $H^{\pm}W^{\mp}$ and $t\bar{t}\ra H^{\pm}W^{\mp}b\bar{b}$ are compared in both final states. A set of simulation packages is used for the analysis and a set of selection cuts is applied on the main distinguishing kinematic variables to increase the signal to background ratio. Finally the signal statistical significance is evaluated as a function of the charged Higgs mass and \tanb~and 5$\sigma$ and 3$\sigma$ contours are plotted.
\\
\section{Signal and Background Processes}
The signal process which is analyzed in this work is $b\bar{b}\ra H^{\pm}W^{\mp}$ followed by the decay $H^{\pm}\ra \tau^{\pm}\nu_{\tau}$. The charged Higgs decay to $t\bar{b}$ is not studied in this work because it is expected to be hard to be distinguished from the large hadronic background although it starts to be the main decay channel with increasing charged Higgs mass in the region above the top quark mass. This argument follows the results quoted in Refs. \cite{LowetteCH,CHATLAS,35}. The charged Higgs decays to $c\bar{s}$, $c\bar{b}$, $\mu^{+}\bar{\nu}$ and supersymmetric states are not studied in this work because they are negligible decays for heavy charged Higgs. For the light charged Higgs these decay channels grow but still remain at least two orders of magnitude smaller than the main decay channel i.e. $H^{+}\ra \tau\nu$ Ref. \cite{LCHCMS}. The first two i.e. $H^{+}\ra c\bar{s}$ and $H^{+}\ra c\bar{b}$, suffer also from the large hadronic final state background. \\
The reason for using only $b\bar{b}$ initiated process is that the dominant contribution comes from $b\bar{b}$ and the relative contribution of $gg$ process becomes small and negligible ($O(10^{-2})$) at high \tanb$~$ which is the region of interest in this study. This argument is in accord with theoretical studies of this process in Refs. \cite{24} and \cite{25}. Depending on the W boson decay, two search categories are defined as the leptonic final state i.e. $W^{\pm}\ra \ell \nu$ where $\ell = e$ or $\mu$ and the hadronic final state i.e. $W^{\pm}\ra jj$ where $j$ is a light quark initiating a jet. Therefore the two signal processes under study are $b\bar{b}\ra H^{\pm}W^{\mp}\ra \tau^{\pm}\nu_{\tau}\ell \nu$ and $b\bar{b}\ra H^{\pm}W^{\mp}\ra \tau^{\pm}\nu_{\tau}jj$.
\\
The background processes which are studied in this analysis are $t\bar{t}$, $WW$ and $W$+jets. In the search for the leptonic final state, the $t\bar{t}$ process is forced to produce the following final states $t\bar{t}\ra WbWb\ra \ell\tau bb E^{miss}_{T}$ or $\ell jj bb E^{miss}_{T}$. In the same search the following final state for $WW$ process is considered: $WW\ra \ell\tau E^{miss}_{T}$ or $\ell jj E^{miss}_{T}$. The $W$+jets is used with $W\ra \ell\nu$. All these processes are included to account for non-identified jets which escape the reconstruction and also the $\tau$ jet fake rate which comes from light quark jets present in the event which pass the $\tau$ identification algorithm.\\
Subsequently the hadronic final state search is performed with forcing $t\bar{t}$ events to decay to the following final states $t\bar{t}\ra WbWb\ra \tau jj bb E^{miss}_{T}$ or $\tau \tau bb E^{miss}_{T}$ or $jjjj bb$ while $WW$ process is forced to decay to 
$WW\ra \tau jj E^{miss}_{T}$ or $\tau \tau E^{miss}_{T}$ or $jjjj$. In the $W$+jets simulation the $W$ boson can decay as $W\ra\tau E^{miss}_{T}$ or $W\ra\ jj$.
\\
\section{Event Simulation}
Signal events are simulated using PYBBWH code Ref. \cite{pybbwh} which is linked to PYTHIA 6.4.21 Ref. \cite{pythia}. The PYBBWH is currently the only Monte Carlo package for generating $b\bar{b}\ra H^{\pm}W^{\mp}$. This is the dominant production process in the high \tanb$~$ region and provides a reasonable estimate of the signal rate. Therefore throughout the paper this process is used and declared as the signal process. As stated in the previous sections, although there can be negative contributions due to the higher order QCD or EW corrections including supersymmetric effects, these effects can be compensated by choosing a proper scenario of the 2HDM paremeters. \\
The $\tau$ leptons polarization and decays are controlled by the TAUOLA package (Refs. \cite{tauola,tauola2,tauola3}) which is linked to PYTHIA. Throughout the analysis $\tau$ leptons are identified through their hadronic decays which results in a narrow $\tau$ jet which will be described in detail later in the next sections. The analysis is based on parton showering and hadronization. The jet reconstruction is performed using PYTHIA built-in jet reconstruction tool, PYCELL, with a cone size of $0.5$. Only jets within the pseudorapidity range of $|\eta|<3.0$ are reconstructed. The parton distribution function used in the analysis is MRST 2004 NNLO which is used by linking LHAPDF 5.8.1 Ref. \cite{lhapdf} to the PYTHIA event generator. For the signal simulation the $m_{h^{0}}-max$ scenario is used with the following parameters: $M_{2}=200$ GeV, $M_{\tilde{g}}=800$ GeV, $\mu=200$ GeV and $M_{SUSY}=1$ TeV.
\\
\section{Signal and Background Cross Sections}
The signal cross section is calculated at tree level by the joint PYTHIA+PYBBWH package using the MRST 2004 NNLO PDF set. The background cross sections are calculated using MCFM 5.7 Ref. \cite{mcfm} at next to leading order with the same PDF set as above. Table \ref{Xsec1} lists the total cross section of the signal with different charged Higgs masses at \tanb=30 and Tab. \ref{Xsec2} shows the calculated cross sections of different background processes used in the analysis.
\begin{table}
\begin{center}
% use packages: array
\begin{tabular}{|c|c|c|c|c|}
\hline
& \multicolumn{4}{|c|}{Signal, $H^{\pm}W^{\mp}$} \\ 
\hline 
$m_{(H^{\pm})}$ & 150 GeV  & 175 GeV & 190 GeV & 200 GeV \\
\hline
Cross Section & 0.15 pb & 0.12 pb & 0.103 pb & 0.095 pb \\
\hline
\end{tabular}
\end{center}
\caption{Signal cross sections with \tanb=30. \label{Xsec1}}
\end{table}
\begin{table}
\begin{center}
% use packages: array
\begin{tabular}{|c|c|c|c|}
\hline
& \multicolumn{3}{|c|}{Background}\\ 
\hline
Process & $W^{+}W^{-}$ & $t\bar{t}$ & $W+$jets\\
\hline
Cross section & 115.5$\pm$0.4 pb & 878.7$\pm$0.5 pb & 187.1$\pm$0.1 nb \\
\hline
\end{tabular}
\end{center}
\caption{Background cross sections used in the analysis. \label{Xsec2}}
\end{table}
The branching ratios used in the analysis are BR$(t\ra W^{\pm}b)=0.98$, BR$(t\ra H^{\pm}b)=0.02$ for $m_{(H^{\pm})}=150$ GeV and \tanb=30, BR$(H^{\pm}\ra \tau \nu)=0.987$ which is almost constant for the mass range $m_{(H^{\pm})}<175$ GeV. The $W$ boson is considered to decay according to the following branching ratios BR$(W^{\pm}\ra \ell \nu)=0.106$ and BR$(W^{\pm}\ra jj)=0.68$ which are taken from Particle Data Group (PDG) at Ref. \cite{pdg}. The branching ratios of charged Higgs decays are calculated using HDECAY 3.51 Ref. \cite{hdecay}. The branching ratio of top quark decay to charged Higgs is taken from PYTHIA. 

\section{Event Analysis and Signal Search}
In this section the event analysis is described in detail. The search is divided into two regions of low mass and high mass charged Higgs with the separating point set to $m_{(H^{\pm})}\simeq 175$ GeV, which is the end point beyond which the production process $t\bar{t}$ is pure SM process and no on-shell $t\ra H^{+}b$ decay is allowed. \\
The basic selection cuts are defined and used depending on the final state under study. In the leptonic final state a signal event contains a muon or an electron, from the $W$ boson decay, with approximately the same kinematic properties, a $\tau$ lepton from the charged Higgs decay and some amount of missing transverse energy originating from both $W$ and $\tau$ decays. Therefore a set of selection requirements is designed to identify these three physical objects in the event taking into account the kinematic differences between the signal and background to increase the signal to background ratio. Similarly a signal event in the hadronic final state contains two jets from the $W$ boson decay, a $\tau$ lepton from the charged Higgs decay and some missing transverse energy mainly from the $\tau$ lepton decay.\\
The kinematic thresholds applied on transverse momenta of muons and electrons as well as the missing transverse energy is proposed by comparing the distributions of these quantities between the signal and background samples. On the other hand the $\tau$ identification algorithm used in the analysis tries to be as close to real LHC algorithms as possible. The leptonic decay of the $\tau$ lepton is not analyzed due to the lower branching ratio of the leptonic decay and the soft lepton $p_{T}$ spectrum from the $\tau$ leptonic decay.\\
In total four points in the parameter space are studied. One point is in the low mass region with $m_{(H^{+})}=150$ GeV. At this point two leptonic and hadronic final states are studied and compared with the main process which is the $t\bar{t}\ra H^{\pm}W^{\mp}b\bar{b}$. The amount of excess one can observe by including the $H^{\pm}W^{\mp}$ signal is then estimated in the low mass region.\\
The other three points i.e. $m_{(H^{+})}=175,~190$ and $200$ GeV lie in the high mass region and are studied each in the leptonic and hadronic final states separately. \\
As will be seen in the next sections, the signal selection efficiency increases with increasing the charged Higgs mass due to the harder kinematics of physical objects in the event. However the cross section decreases more rapidly than the increasing rate of the selection efficiency, thus demanding higher \tanb$~$ values to compensate the net decrease in the signal statistics when higher charged Higgs masses are studied. Therefore a $5\sigma$ and $3\sigma$ contour is plotted to present the accessible regions of the parameter space. Using these plots, a conclusion about higher charged Higgs masses can easily be made by extrapolating the contour. However the main conclusion is that the most sensitive part of the parameter space is the region studied in this analysis and other parts have less sensitivity and with increasing charged Higgs mass, higher \tanb$~$ values are required to reach a $5\sigma$ discovery.

\subsection{Heavy Charged Higgs Search, Leptonic Final State}
\begin{figure}
\begin{center}
\includegraphics[width=0.80\textwidth]{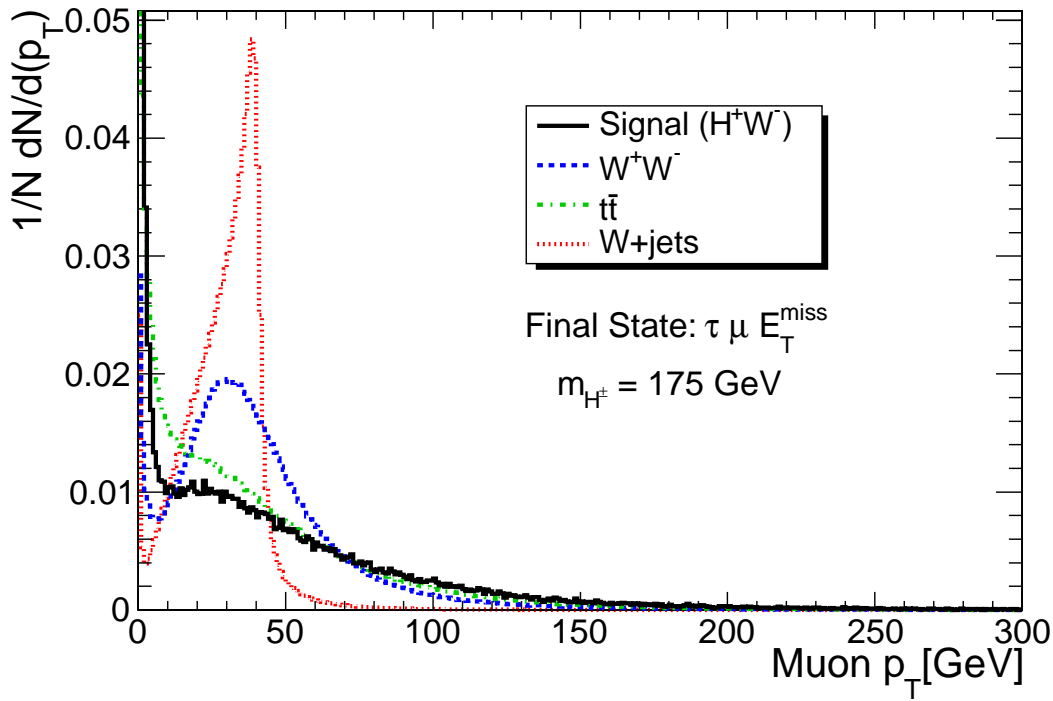}
\end{center}
\caption{Transverse momentum distribution of muons in signal and background events in the $\tau\mu E^{miss}_{T}$ final state .\label{mupt}}
\end{figure}

\begin{figure}
\begin{center}
\includegraphics[width=0.80\textwidth]{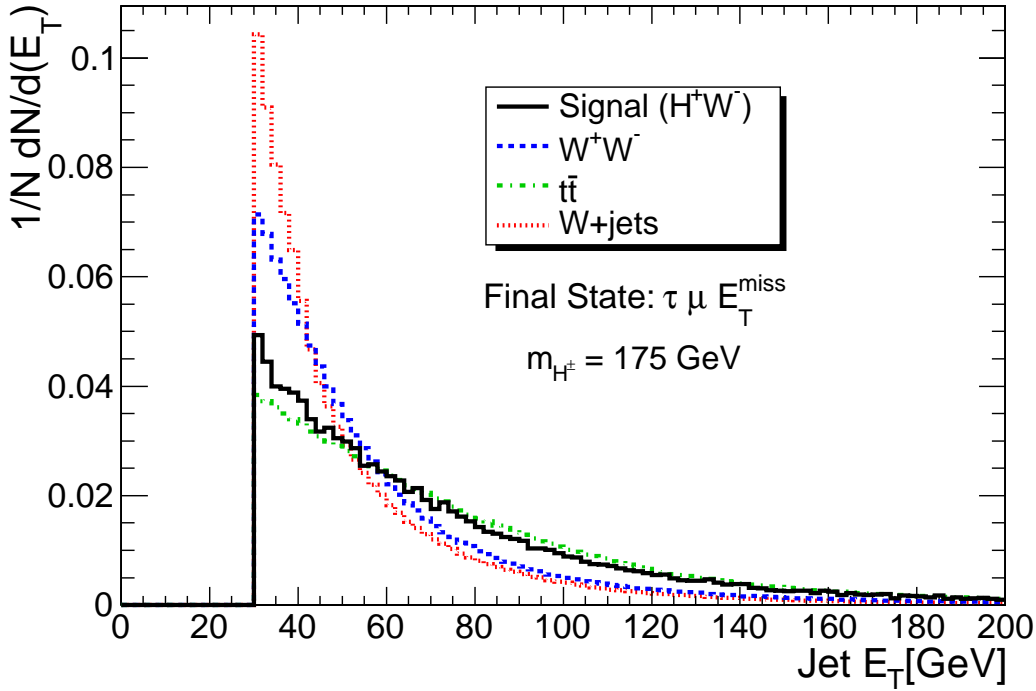}
\end{center}
\caption{Transverse energy distribution of reconstructed jets in signal and background events in the $\tau\mu E^{miss}_{T}$ final state. All jets with $E_{T} < 30$ GeV are thrown away in the event as a general cut-off on the jet transverse energy. This is adopted usually in full simulation analyses to reject contribution of soft jets from pile-up events and also avoid reliability issues related to performance of the jet reconstruction algorithm at low transverse energies. \label{jetet}}
\end{figure}

\begin{figure}
\begin{center}
\includegraphics[width=0.80\textwidth]{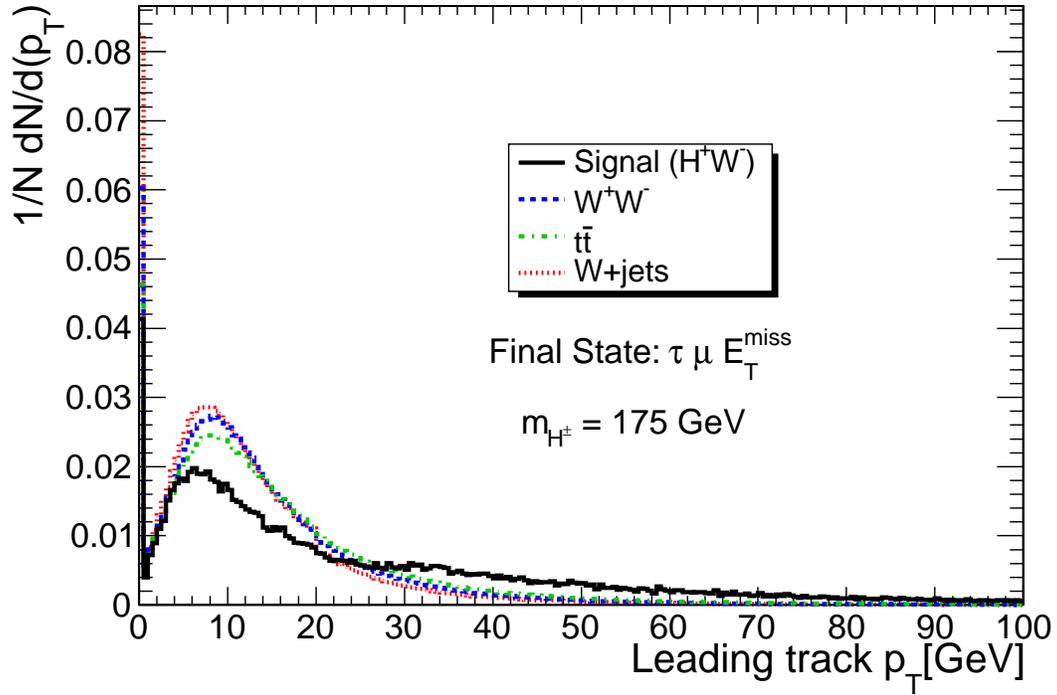}
\end{center}
\caption{Transverse momentum distribution of the leading track in the $\tau$ jet cone.\label{ltpt}}
\end{figure}

In the following, a heavy charged Higgs with a mass of 175 GeV is analyzed in leptonic final state. This is a point in parameter space close to the defined border at which the on-shell $t\bar{t}\ra H^{\pm}W^{\mp}b\bar{b}$  process dies and the charged Higgs may only be produced by $gg\ra t\bar{b}H^{-}+gb\ra tH^{-}$ or $H^{\pm}W^{\mp}$. The aim of this work is to solely estimate the amount of excess one may observe over SM processes if a search is designed for $H^{\pm}W^{\mp}$ events. The set of selection cuts is described below for $\tau\mu E^{miss}_{T}$ final state, however a similar contribution from the electronic decay of the W boson i.e. $\tau e E^{miss}_{T}$ final state is expected. This is based on the assumption that in high energy events produced at LHC environment the mass difference between muons and electrons is negligible thus expecting the same kinematics between muonic and electronic final states and a reasonable lepton reconstruction efficiency achieved at LHC detectors ensures that no sizable lepton loss is expected. Discussion about the lepton isolation efficiencies and the performance of jet reconstruction algorithms in events containing electrons is beyond the scope of this work and needs a reasonable detector simulation. In the following event selection cuts are described for the final state containing muons, however, the signal significance is calculated assuming the same contribution from the final state containing electrons. An event must contain all the following objects to be selected:
\\
\textbullet One muon satisfying Eq. \ref{mukin}:
\begin{equation}
p^{\mu}_{T}> 50~{\rm GeV},~ |\eta|<2.5 \ .
\label{mukin}
\end{equation}
where $\eta$ is pseudorapidity defined as $$\eta\equiv -\log\left[ \tan \frac{\theta}{2} \right]$$
in which $\theta$ is the angle between the momentum and the beam axis.
The adopted threshold in Eq. \ref{mukin} has been inspired by Fig. \ref{mupt}.
\\
\textbullet One reconstructed jet satisfying Eq. \ref{taukin}:
\begin{equation}
 E^{\textnormal{jet}}_{T}>50 ~ \textnormal{GeV}, ~ |\eta|<2.5 \ .
\label{taukin}
\end{equation}
This cut is also adopted in accord with Fig. \ref{jetet}. 
To reject the contribution of non-isolated muons which come mainly from heavy meson decays, the selected jet and muons with with $p_{T}>20$ GeV should be separated enough with the following requirement:
\begin{equation}
\Delta R_{(\textnormal{jet,$\mu$})}>0.4 \ 
\label{isol}
\end{equation}
where $\Delta R$ is defined as $\Delta R= \sqrt{(\Delta \eta)^2+(\Delta \phi)^2}$.\\
The $\tau$ identification algorithm is then applied on the reconstructed jet in the event. A jet is accepted if it passes all the requirements of $\tau$-id. Although in a full simulation analysis there may be sophisticated requirements for $\tau$ jets, here the basic requirements are applied. These requirements identify $\tau$ jets with a reasonable purity. However the $\tau$ fake rate (contribution of non-$\tau$ jets which pass the $\tau$-id algorithm) may be higher than in a real analysis with full $\tau$-id cuts. This fact arises hopes for better results with less background contamination in the signal region. Verification of this issue is beyond the scope of this analysis which is providing prospects of the observability of this channel at the LHC and motivation of performing this analysis by LHC experiments, CMS and ATLAS.\\
The $\tau$-id which is used in the analysis is similar to the one used by the CMS collaboration Ref. \cite{tauid}. 
In this analysis all decays of $\tau$ leptons are turned on but the search is only based on the hadronic decays which in total account for $\sim 65\%$ of the total decay width of the $\tau$ lepton. The $\tau$ lepton in its hadronic decay produces predominantly one or three charged pions. Due to the low charged track multiplicity in the final state, the charged tracks (pions) in the $\tau$ hadronic decay acquire relatively a higher transverse momentum compared to tracks of light quark jets. To study this effect, a jet-track matching cone of $\Delta R=0.1$ is considered around the jet axis. The hardest charged track in the matching cone is considered as the charged pion from the $\tau$ lepton decay. Figure \ref{ltpt} shows distribution of the leading track transverse momentum in the matching cone around the jet. To exploit this feature a transverse momentum threshold is applied as the following:
\begin{equation}
 p^{\textnormal{leading track}}_{T}>20 ~ \textnormal{GeV} \ .
\end{equation}

\begin{figure}
\begin{center}
\includegraphics[width=0.80\textwidth]{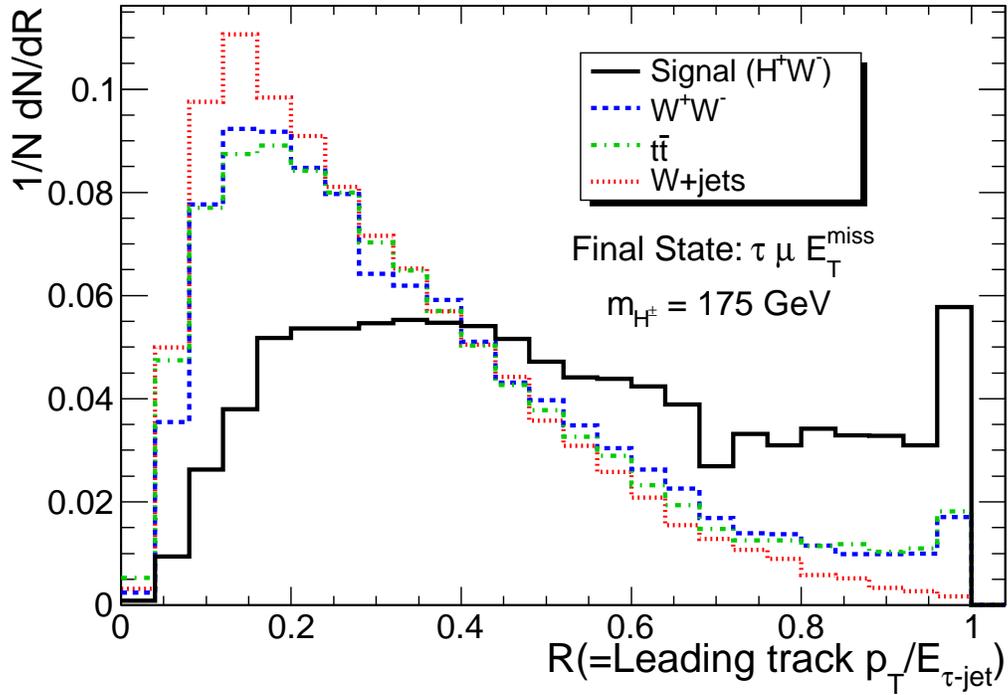}
\end{center}
\caption{Distribution of the leading track $p_T$ divided by the $\tau$ jet energy.\label{R}}
\end{figure}

\begin{figure}
\begin{center}
\includegraphics[width=0.80\textwidth]{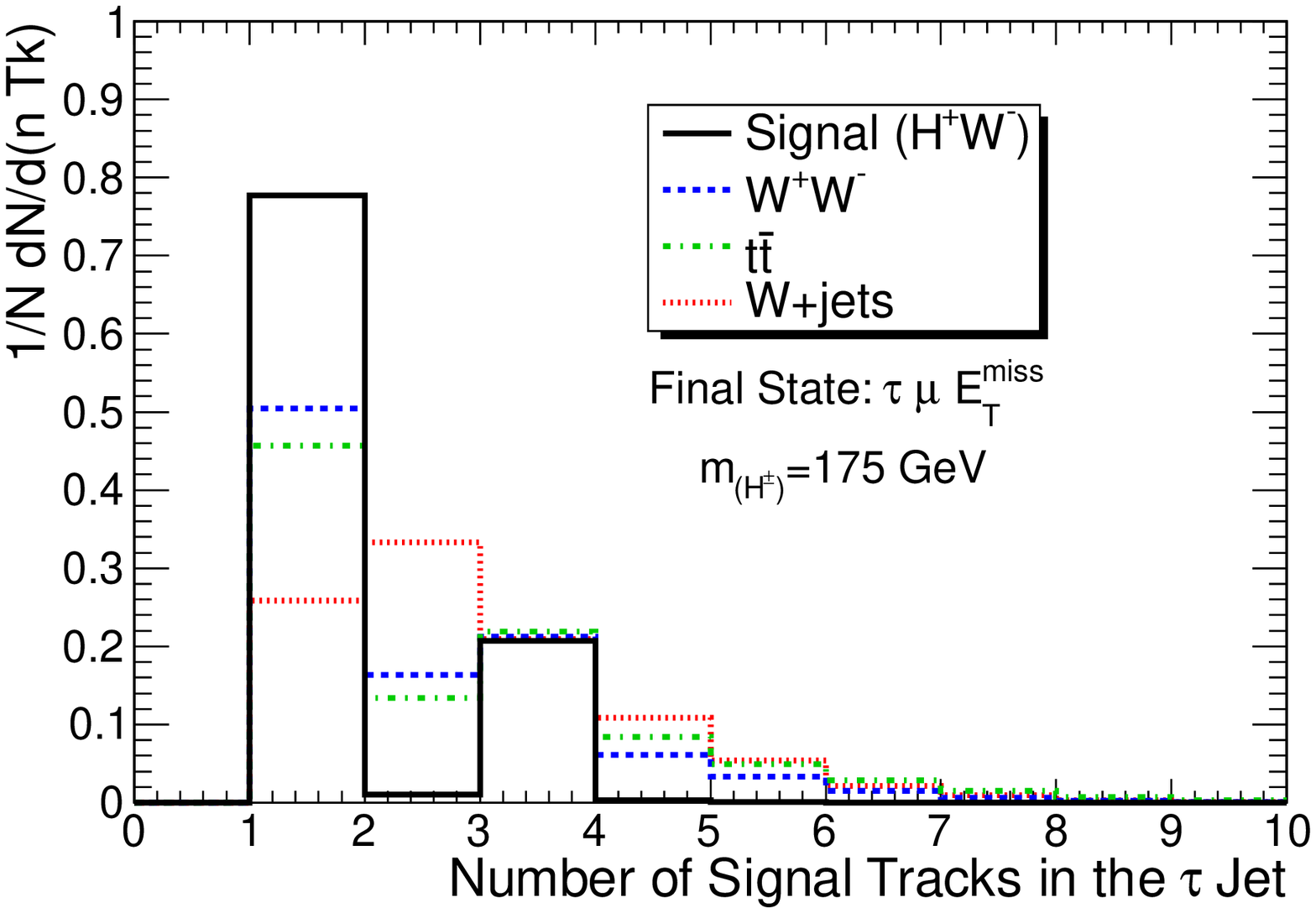}
\end{center}
\caption{Number of tracks in the signal cone around the leading track of the $\tau$ jet.\label{nst}}
\end{figure}

\begin{figure}
\begin{center}
\includegraphics[width=0.80\textwidth]{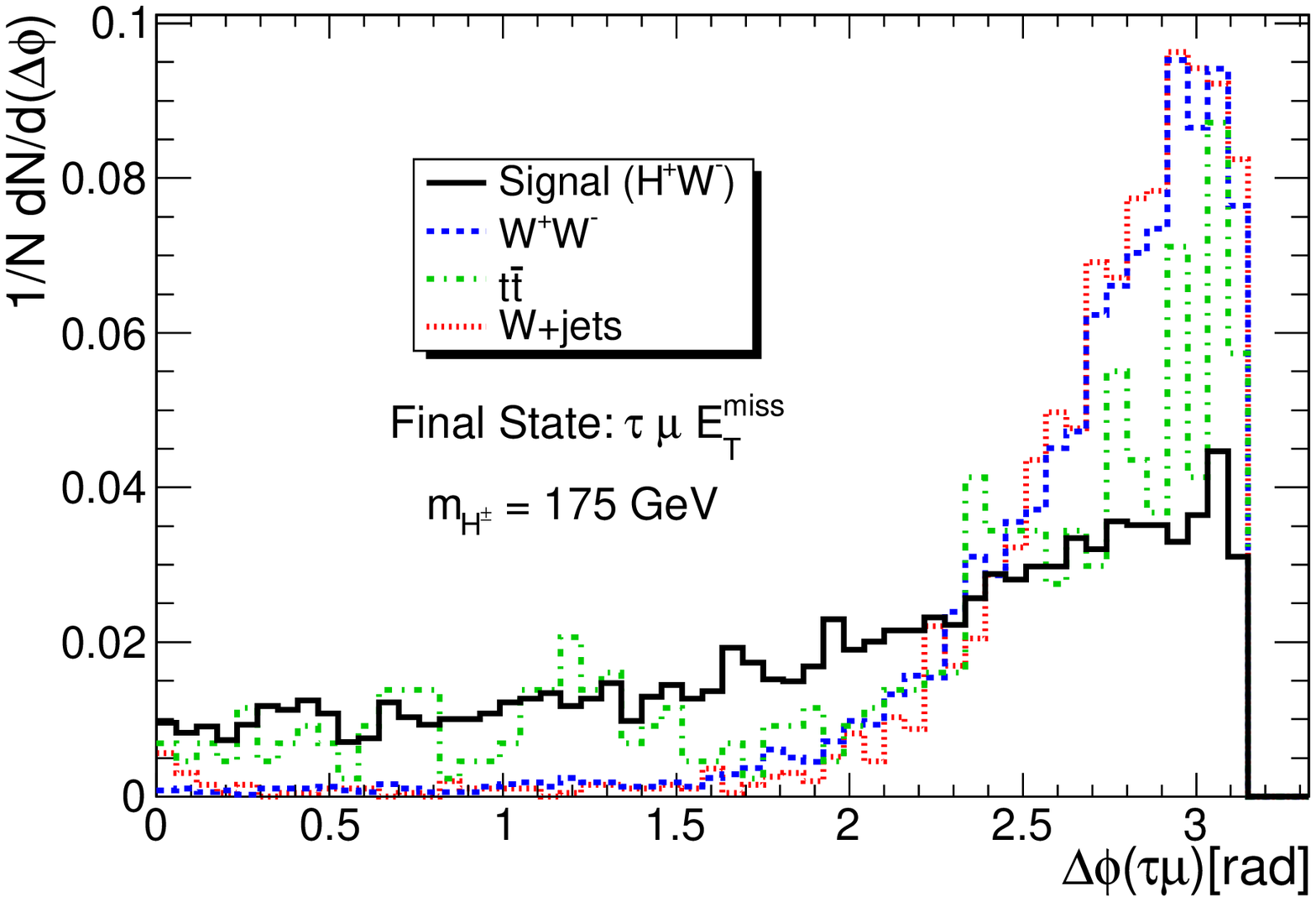}
\end{center}
\caption{The azimuthal angle between the muon and the $\tau$ jet candidate in the $\tau\mu E^{miss}_{T}$ final state.\label{dphi}}
\end{figure}

\begin{figure}
\begin{center}
\includegraphics[width=0.80\textwidth]{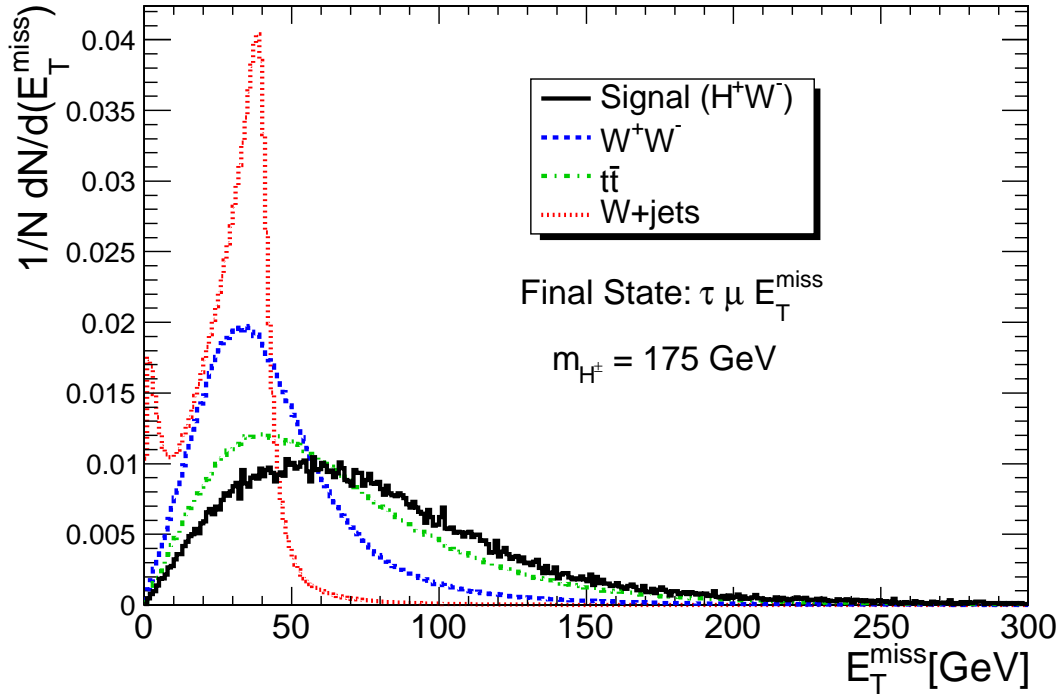}
\end{center}
\caption{Missing transverse energy distribution in signal and background events in the $\tau\mu E^{miss}_{T}$ final state. \label{met}}
\end{figure}

Since $\tau$ jets consist of few charged tracks, they are considered as isolated jets. To check if a $\tau$ jet is isolated an isolation cone and a signal cone is defined respectively with a cone size of $\Delta R<0.4$ and $\Delta R<0.07$ around the leading track. The isolation is applied by requiring no charged track with $p_{T}>1$ GeV to be in the isolation annulus defined as $0.07<\Delta R<0.4$.
The low charged track multiplicity in the $\tau$ jets also implies that the leading track in the jet cone carries a larger fraction of the $\tau$ jet energy compared to quark jets in background events. This effect can be verified by plotting distribution of the leading track $p_{T}$ divided by the $\tau$ jet energy as shown in Fig. \ref{R}.
In order to keep the signal statistics at a reasonable amount, a soft cut on this quantity is applied as the following:
\begin{equation}
\textnormal{R}=\textnormal{leading~track}~p_{T}~/~E_{\tau-\textnormal{jet}}>0.2 \ .
\end{equation}
The number of charged tracks in the $\tau$ jet is also evaluated by a search in the signal cone and the following requirement is applied:
\begin{equation}
\textnormal{Number of signal tracks} = 1 ~ \textnormal{or} ~ 3 \ .
\end{equation}
Figure \ref{nst} shows distribution of the number of signal tracks in signal events before applying the cut.
As seen from Fig. \ref{nst}, $\tau$ jets have undergone 1- or 3-prong decays predominantly.
The azimuthal angle between the muons and $\tau$ leptons is also investigated as shown in Fig. \ref{dphi}.
As is observed from Fig. \ref{dphi}, background events tend to produce a harder back-to-back topology, however, in order to keep the signal statistics, no cut on this kinematic variable is applied. In the next step the $\tau$ lepton charge is calculated as the sum of charges of tracks in the signal cone. Since muons and $\tau$ leptons in signal events are produced with opposite charges, the following requirement is applied:
\begin{equation}
\textnormal{Muon~ charge +~} \tau \textnormal{~jet~ charge} =0 \ .
\end{equation}
\textbullet A reasonable amount of missing transverse energy should be left in the event. This is because of spin considerations and the fact that in $H^{\pm}W^{\mp}$ events, neutrinos tend to fly collinearly which is the result of charged Higgs boson being a spinless particle. The transverse missing energy distribution is studied as shown in Fig. \ref{met}. As seen from Fig \ref{met}, the main backgrounds are all produced with a low $E^{miss}_{T}$ compared to signal events.
The following requirement is therefore applied on $E^{miss}_{T}$:
\begin{equation}
E^{miss}_{T}>50~ \textnormal{GeV} \ .
\end{equation}

In order to have an estimation of the signal significance, all selection cuts are applied sequentially and their relative efficiencies with respect to the previous cut and the number of events which survive after each cut are listed in special tables. Table \ref{seleff1} lists selection cuts efficiencies and the number of signal and background events remaining after each cut.
\begin{table}
\begin{center}
% use packages: array
\begin{tabular}{|l|c|c|c|c|}
\hline
Process & $H^{\pm}W^{\mp}$ & $W^{+}W^{-}$ & $t\bar{t}$ & W+jets \\
\hline
Total cross section [pb] & 0.12 & 115.5 & 879 & 187100      \\
\hline
Number of events at 30 $fb^{-1}$ & 369 & 577577 & 4394086 & 5.9$\times 10^{8}$    \\
\hline
N Muons = 1 & 172(46.7$\%$) & 141662(24.5$\%$) & 1.73e+06(39.3$\%$)  &  1.34e+07(2.3$\%$)   \\
\hline
N Jets = 1 & 76(44.2$\%$) & 62159(43.9$\%$) & 120121(6.9$\%$) & 6.2e+06(46.6$\%$)    \\
\hline
leading track $p_{T} > 20$ GeV & 43(56.5$\%$) & 19327(31.1$\%$) & 35557(29.6$\%$) & 1.5e+06(24.9$\%$)  \\
\hline
Isolation & 35(81.7$\%$) & 3676(19$\%$) & 3247(9.1$\%$)  &  133281(8.6$\%$)   \\
\hline
R $>$ 0.2 & 31(87.4$\%$) & 2617(71.2$\%$)  & 2223(68.5$\%$) & 86887(65.2$\%$)    \\
\hline
1- or 3-prong decay & 30(98.4$\%$) & 2177(83.2$\%$) & 1916(86.2$\%$)  & 38389(44.2$\%$) \\
\hline
Opposite charge & 30(99.8$\%$) & 2146(98.6$\%$) & 1885(98.4$\%$)   &  35577(92.6742$\%$)   \\
\hline
$E^{miss}_{T} > 50$ GeV & 21(71.2$\%$) & 776(36.1$\%$)& 1112(59$\%$) & 3737(10.5$\%$)    \\
\hline
Total efficiency & 5.83$\%$ & 0.13$\%$ & 0.025$\%$  &  6.3$\times 10^{-4}\%$     \\
\hline
Expected events at $30fb^{-1}$ & 21 & 776 & 1112  &  3737   \\
\hline
\end{tabular}
\end{center}
\caption{Selection efficiencies and remaining number of signal and background
events after each cut in the $\tau\mu E^{miss}_{T}$ final state. The charged Higgs mass is set to 175 GeV and \tanb = 30. Numbers in parentheses are relative efficiencies in percent
with respect to the previous cut. Branching ratios have been taken into account in transition from the second to third row.\label{seleff1}}
\end{table}
The number of signal and background events remaining after all selection cuts (the last row of Tab. \ref{seleff1} multiplied by a factor of two to account for electrons) is used to calculate the signal significance according to Eq. \ref{signif}.
\begin{equation}
\textnormal{Signal~ Significance}=\frac{N_{S}}{\sqrt{N_{B}}}
\label{signif}
\end{equation}
Using Eq. \ref{signif} the signal significance with $m_{(H^{\pm})}=175$ GeV and \tanb=30 is estimated to be $\sim 0.4\sigma$ with a data corresponding to 30 \invfb~integrated luminosity. At high luminosity run of LHC when 300\invfb~data is collected the signal significance turns out to be $1.3\sigma$ with $m_{(H^{\pm})}=175$ GeV and \tanb=30. The border of $5\sigma$ contour (the point below which the signal significance is below $5\sigma$) starts from \tanb$ \sim 60$ and goes to higher \tanb values for heavier charged Higgs bosons. As a result a charged Higgs with a mass $m_{(H^{\pm})}=175$ GeV is observable in the leptonic final state search if $tan\beta>60$ and lower \tanb~values are out of $5\sigma$ reach.\\
It should be mentioned that using state of the art algorithms developed by LHC experiments, the fake rate is expected to be much smaller than what is observed in this analysis. Since $W+$jets background is the main background and is a source of $\tau$ fake rate, it is expected that this background would be under control in an LHC experiment analysis and lower \tanb~values would be in the $5\sigma$ reach.
 
\subsection{Heavy Charged Higgs Search, Hadronic Final State}
\begin{figure}
\begin{center}
\includegraphics[width=0.80\textwidth]{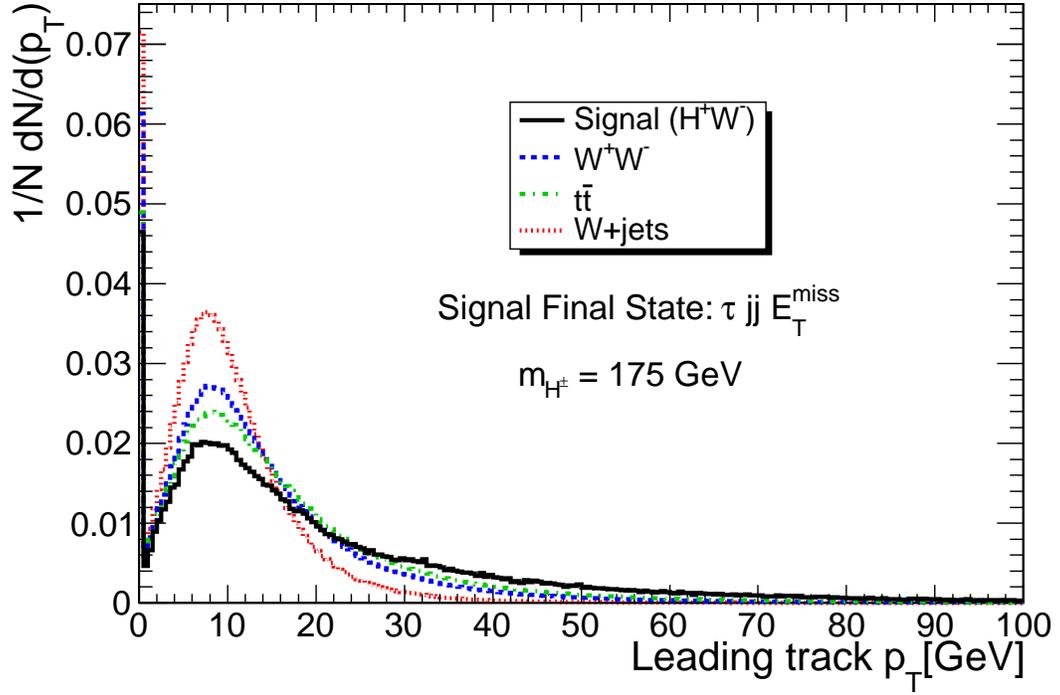}
\end{center}
\caption{Transverse momentum distribution of the leading track in the $\tau$ jet cone.\label{ltpthad}}
\end{figure}

\begin{figure}
\begin{center}
\includegraphics[width=0.80\textwidth]{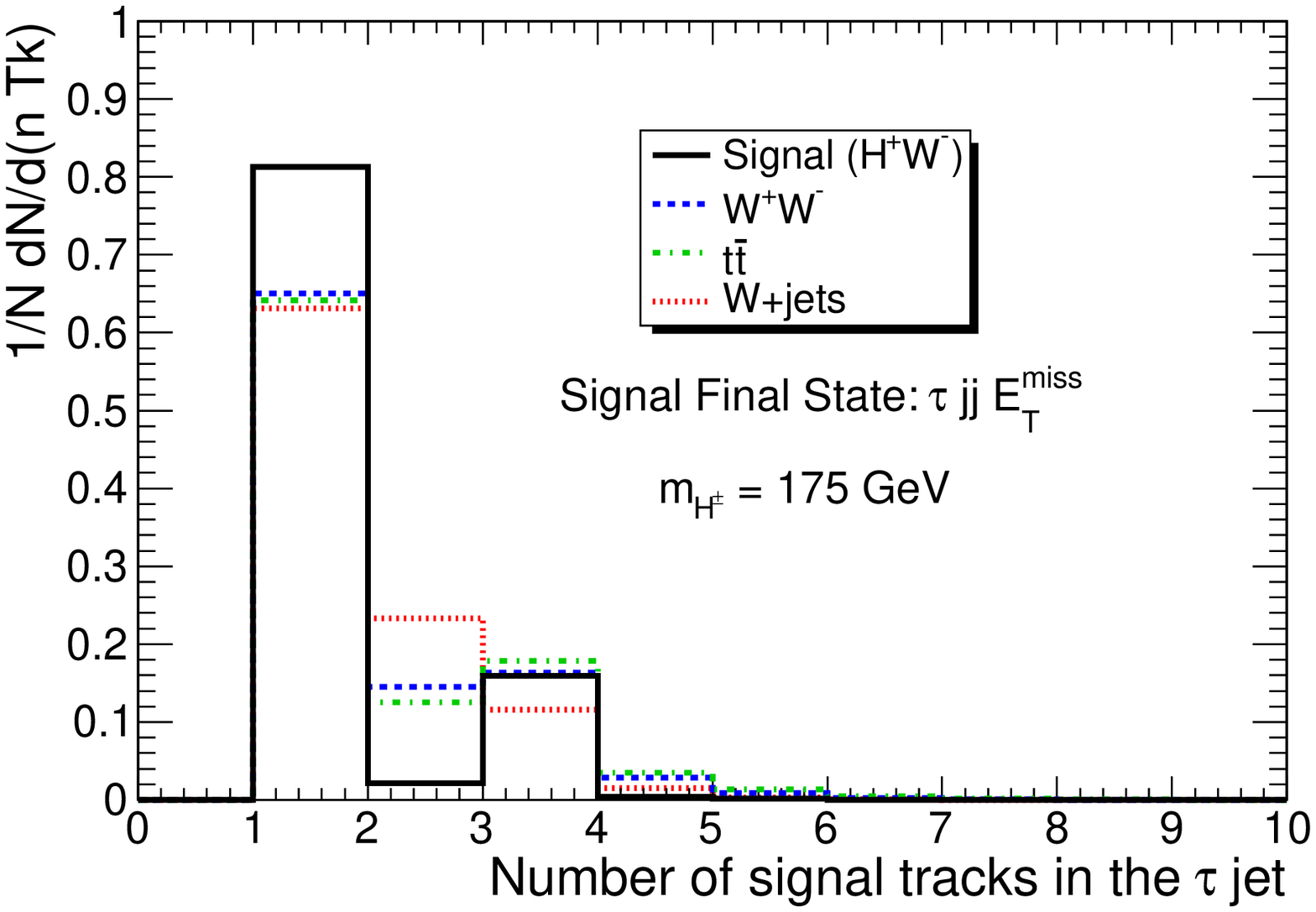}
\end{center}
\caption{Number of tracks in the signal cone around the leading track of the $\tau$ jet.\label{nstkhad}}
\end{figure}

\begin{figure}
\begin{center}
\includegraphics[width=0.80\textwidth]{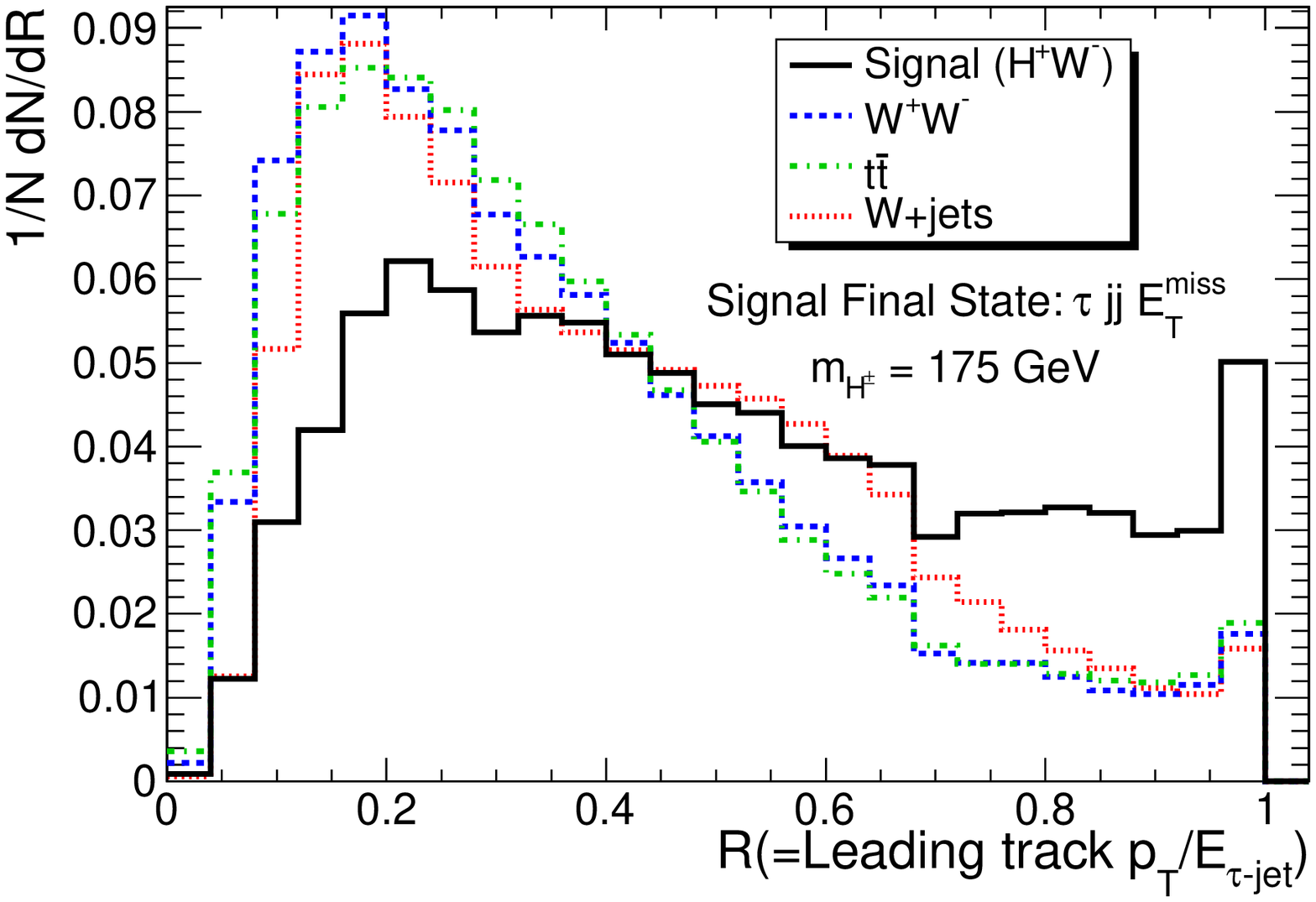}
\end{center}
\caption{Distribution of the leading track $p_T$ divided by the $\tau$ jet energy.\label{Rhad}}
\end{figure}

\begin{figure}
\begin{center}
\includegraphics[width=0.80\textwidth]{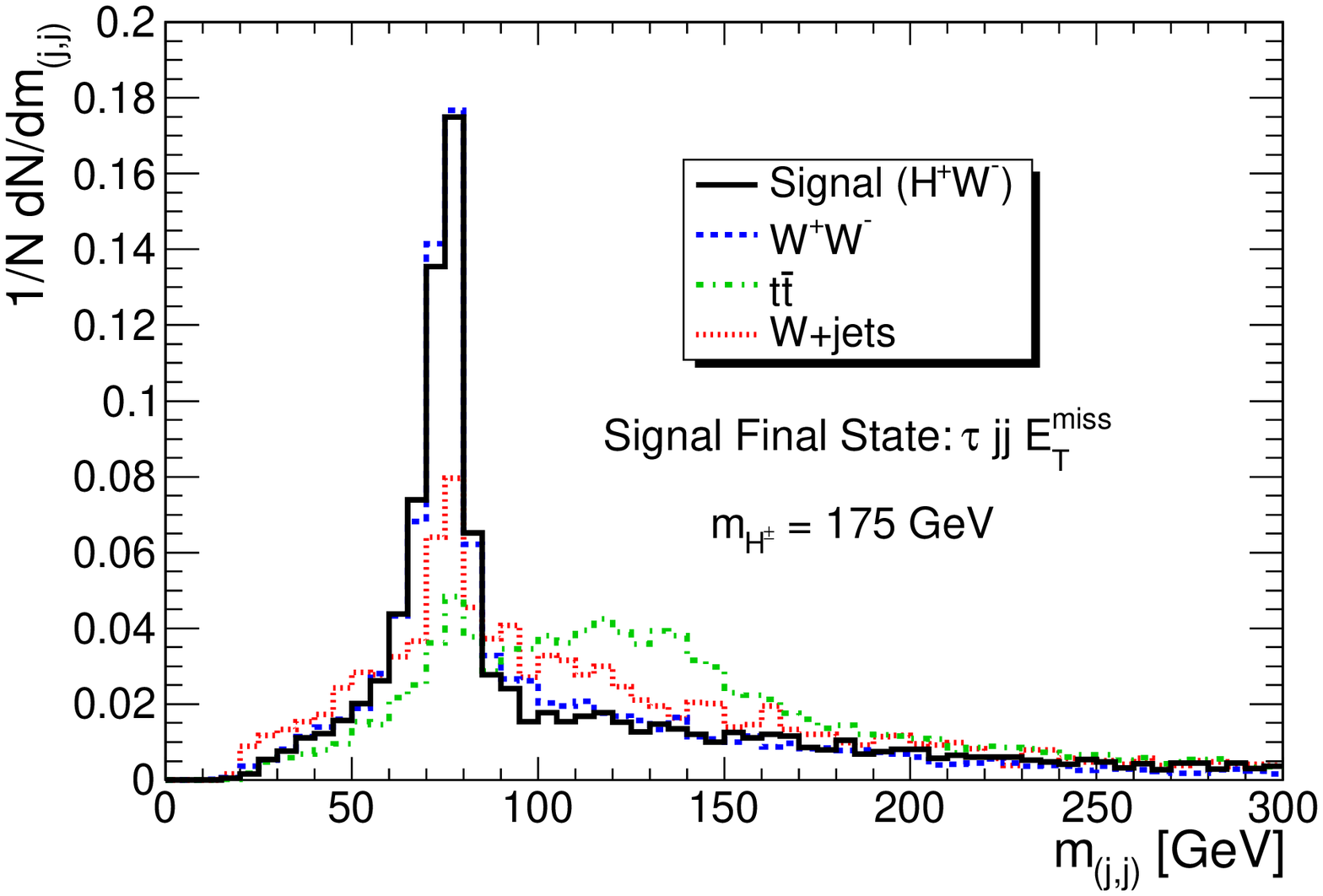}
\end{center}
\caption{Invariant mass of the two jets as the candidate pair from W boson decay. \label{winvmass}}
\end{figure}

\begin{figure}
\begin{center}
\includegraphics[width=0.80\textwidth]{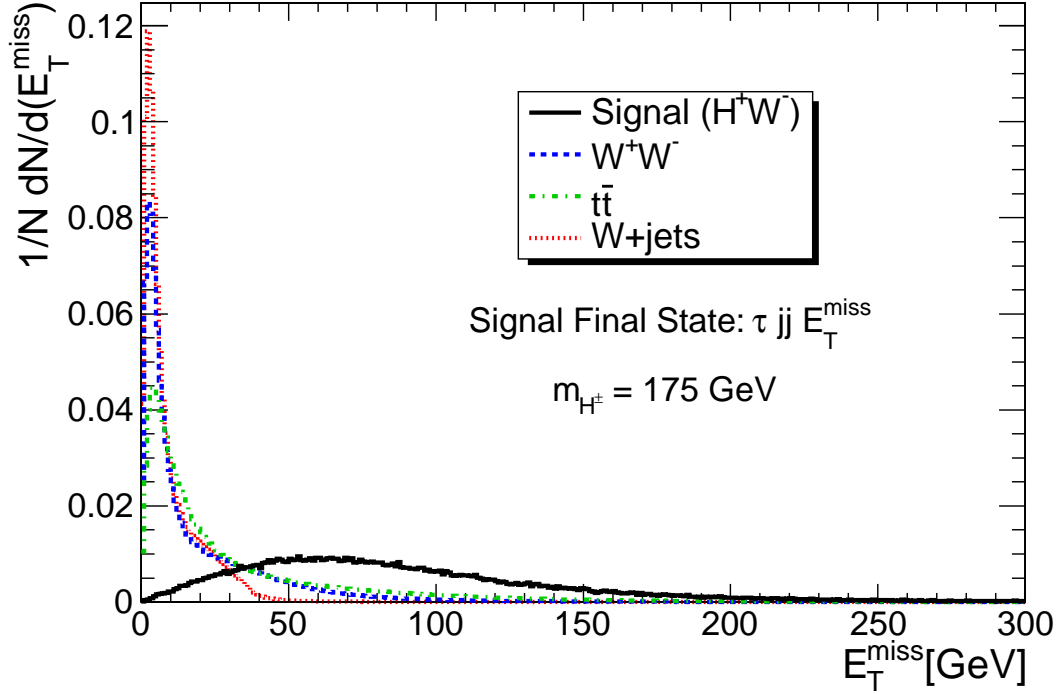}
\end{center}
\caption{Missing transverse energy distribution in signal and background events in the $\tau\ jj E^{miss}_{T}$ final state search. \label{methad}}
\end{figure}

\begin{figure}
\begin{center}
\includegraphics[width=0.80\textwidth]{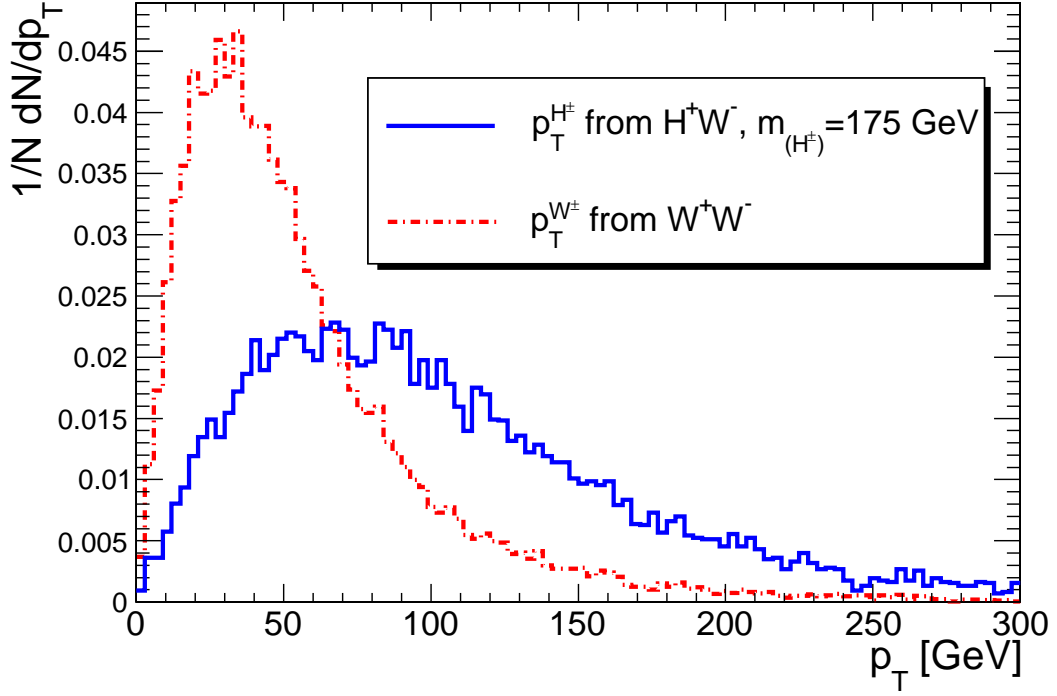}
\end{center}
\caption{The transverse momentum distribution of charged Higgs with $m_{(H^{\pm})}=175$ GeV in $H^{\pm}W^{\mp}$ events and $W$ boson in $W^{+}W^{-}$.\label{chw}}
\end{figure}

\begin{figure}
\begin{center}
\includegraphics[width=0.80\textwidth]{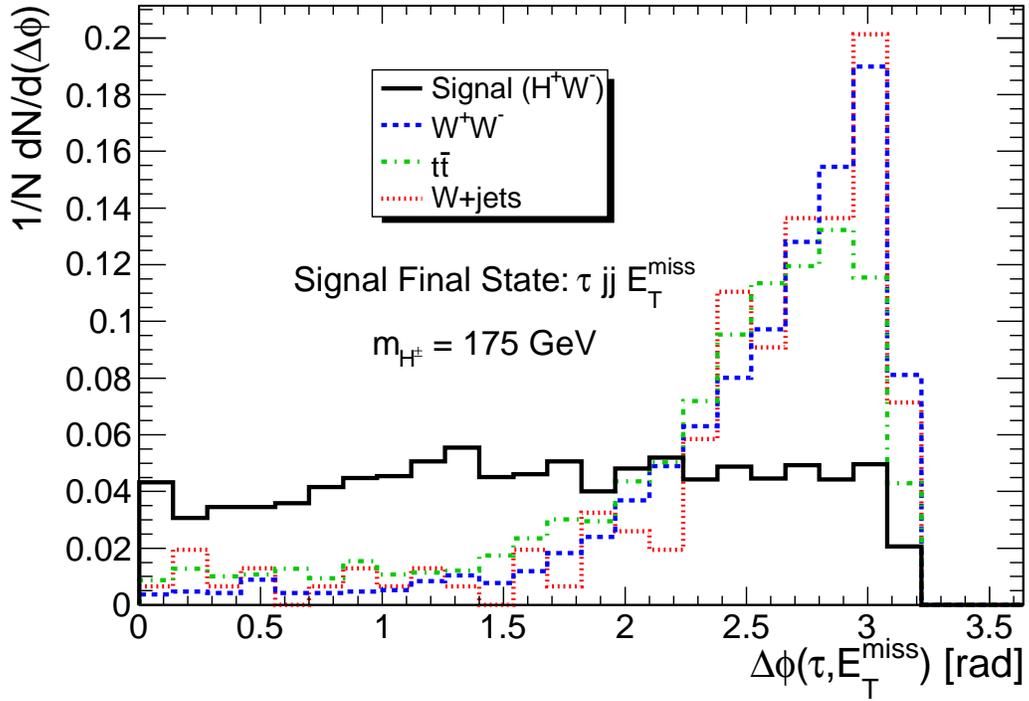}
\end{center}
\caption{The azimuthal angle between the $\tau$ jet and $E^{miss}_{T}$ in the $\tau\ jj E^{miss}_{T}$ final state search. \label{dphihad}}
\end{figure}

The heavy charged Higgs has been studied with this channel in hadronic final state in Ref. \cite{36,37,38}. In this analysis the intention is a closer look to this final state including main background samples and a $\tau$-id as close to what is in use by LHC experiments as possible. The event selection is described in the following. An event must satisfy all the following requirements to be selected:
\\
\textbullet There should be three jets satisfying Eq. \ref{jetkin}. The transverse energy threshold is set to the softer value of 30 GeV compared to the case of leptonic final state in order to avoid unwanted decrease of signal statistics when three jets are required.  
\begin{equation}
E^{\textnormal{jet}}_{T}>30 ~ \textnormal{GeV}, ~~ |\eta|<2.5 \ .
\label{jetkin}
\end{equation}
There has to be exactly three jets in the event and all pairs of jets are required to be well separated by Eq. \ref{sep}.
\be
\Delta R_{(j,j)} > 0.4,~~ \Delta R_{(\tau,j)} > 0.4 \ .
\label{sep}
\ee 
\\
\textbullet The $\tau$-id algorithm is then applied on selected jets in the same way as is done in the leptonic final state search. Exactly one of jets must pass the $\tau$-id algorithm. Figures \ref{ltpthad}, \ref{nstkhad}, \ref{Rhad} show the distributions of related variables used in the $\tau$-id algorithm. 
\\
\textbullet The other two jets remaining in the event are used for $W$ mass reconstruction and their invariant mass is calculated as the $W$ boson mass. The reconstructed $W$ mass should fall within the $W$ mass window defined in Eq. \ref{wmass}.
\be
|m_{(j,j)}-75.| < 20 ~\textnormal{GeV} \ .
\label{wmass}
\ee
The adopted central value for the $W$ invariant mass is a slightly lower than the current values. This is because the reconstructed jet energy may be different from the true value and the calculated invariant mass turns out to be slightly lower than the nominal value of 80 GeV. This is a known issue which is resolved by jet energy correction algorithms developed in LHC experiments. Figure \ref{winvmass} shows the invariant mass of the two jets as the W boson mass candidate. 
\textbullet The missing transverse energy is also used to discriminate between the signal and background. Fig. \ref{methad} shows the distribution of $E^{miss}_{T}$ in signal and background events. As seen from Fig. \ref{methad} the $E^{miss}_{T}$ distribution is much harder in signal events. Since the charged Higgs ($H^{+}$) is spinless, in its decay, it produces a left-handed $\nu_{\tau}$ and a left-handed $\tau^{+}$. In the subsequent $\tau^{+}$ decay, the right-handed $\bar{\nu_{\tau}}$ is kicked back to conserve the mother $\tau^{+}$ helicity. Therefore the two neutrinos preferably fly in the same direction producing a large amount of $E^{miss}_{T}$ in the signal event. The following requirement on $E^{miss}_{T}$ threshold is applied:
\be
E^{miss}_{T} > 50 ~\textnormal{GeV} \ .
\ee
\textbullet A study of azimuthal angle between the $\tau$ jet and $E^{miss}_{T}$ shows that background events tend to produce the neutrino back-to-back with respect to the $\tau$ jet. The main source of $E^{miss}_{T}$ in background events is the neutrino from the $W$ boson decay when $W\ra \tau\nu$ occurs. This decay prefers the $\tau$ jet and the neutrino to fly back-to-back in the $W$ rest frame. The change of the angle between the decay products when transferred to the lab frame is less than the case of charged Higgs boson and thus majority of background events tend to produce high values of $\Delta\phi_{(\tau,E^{miss}_{T})}$. Since the charged Higgs is heavy ($\sim 2 \times m_{W}$), when it is produced, it acquires a higher $p_{T}$ compared to $W$ boson (e.g. from $W^{+}W^{-}$ events) as shown in Fig. \ref{chw} and thus gives a higher boost to the decay products leading to smaller angles in the lab frame. The difference is, of course, more obvious for heavier charged Higgs bosons. This feature results in lower $\Delta\phi_{(\tau,E^{miss}_{T})}$ in signal events as shown in Fig. \ref{dphihad} and is used by applying the requirement as in Eq. \ref{dphicut}.
\be
\Delta\phi_{(\tau,E^{miss}_{T})} < 2.5 \ .
\label{dphicut}
\ee
With the above selection cuts, Tab. \ref{seleff2} lists the selection efficiencies and the remaining number of events after each cut.
\begin{table}
\begin{center}
% use packages: array
\begin{tabular}{|l|c|c|c|c|}
\hline
Process & $H^{\pm}W^{\mp}$ & $W^{+}W^{-}$ & $t\bar{t}$ & W+jets \\
\hline
Total cross section [pb] & 0.12 & 115.5 & 879 & 187100      \\
\hline
Number of events at 30 $fb^{-1}$ & 2363 & 2154302 & 16393214 & 4.4$\times 10^{9}$    \\
\hline
N Jets = 3 & 652(27.6$\%$) & 403831(18.7$\%$) &  3.2e+06(19.5$\%$) & 4e+07(0.9$\%$)  \\
\hline
leading track $p_{T} > 20$ GeV & 509(78$\%$)  & 195290(48.3$\%$) & 2e+06(59.5$\%$) & 1.8e+07(45.5$\%$)   \\
\hline
Isolation & 318(62.5$\%$) & 33839(17.3$\%$) & 244890(12.9$\%$) & 2.4e+06(13.2$\%$)\\
\hline
R $>$ 0.2 &  189(59.4$\%$)  &  12640(37.3$\%$) & 95490(39$\%$) &752987(31.6$\%$)\\
\hline
1- or 3-prong decay & 185(97.8$\%$)  & 10104(80$\%$) & 83556(87.5$\%$) & 498813(66.2$\%$)\\
\hline
Exactly one $\tau$ jet &  184(99.6$\%$) & 10037(99.3$\%$) & 83015(99.3$\%$) & 497493(99.7$\%$)\\
\hline
$W$ mass window &  98(53$\%$)  & 5543(55.2$\%$) & 18803(22.6$\%$) & 169107(34$\%$)\\ 
\hline
$E^{miss}_{T} > 50$ GeV & 73(75.1$\%$)  & 2070(37.3$\%$) & 12205(65$\%$) & 22587(13.3$\%$) \\
\hline
$\Delta\phi_{(\tau,E^{miss}_{T})}$    & 57(78.2$\%$)  & 697(33.7$\%$) & 5615(46$\%$) & 7773(34.4$\%$)\\
\hline
Total efficiency &  2.43$\%$ &  0.032$\%$ & 0.034$\%$ & 0.000177$\%$  \\
\hline
Expected events at $30fb^{-1}$ & 57 &  697 & 5615 & 7773\\
\hline
\end{tabular}
\end{center}
\caption{Selection efficiencies and remaining number of signal and background
events after each cut in the $\tau jj E^{miss}_{T}$ final state search. The charged Higgs mass is set to 175 GeV and \tanb = 30. Numbers in parentheses are relative efficiencies in percent with respect to the previous cut. Branching ratios have been taken into account in transition from the second to third row.\label{seleff2}}
\end{table}
Using the number of events as in Tab. \ref{seleff2} the signal significance is estimated to be $0.5\sigma$ for $m_{(H^{\pm})}=175$ GeV and $tan\beta=30$ at $30~fb^{-1}$ integrated luminosity. At high luminosity run of LHC, the signal significance could increase to $1.5\sigma$. The \tanb~ value for which the signal significance at high luminosity is $5\sigma$ is estimated to be $tan\beta \simeq 55$.
\subsection{Light Charged Higgs Search, Leptonic Final State}
\begin{figure}
\begin{center}
\includegraphics[width=0.8\textwidth]{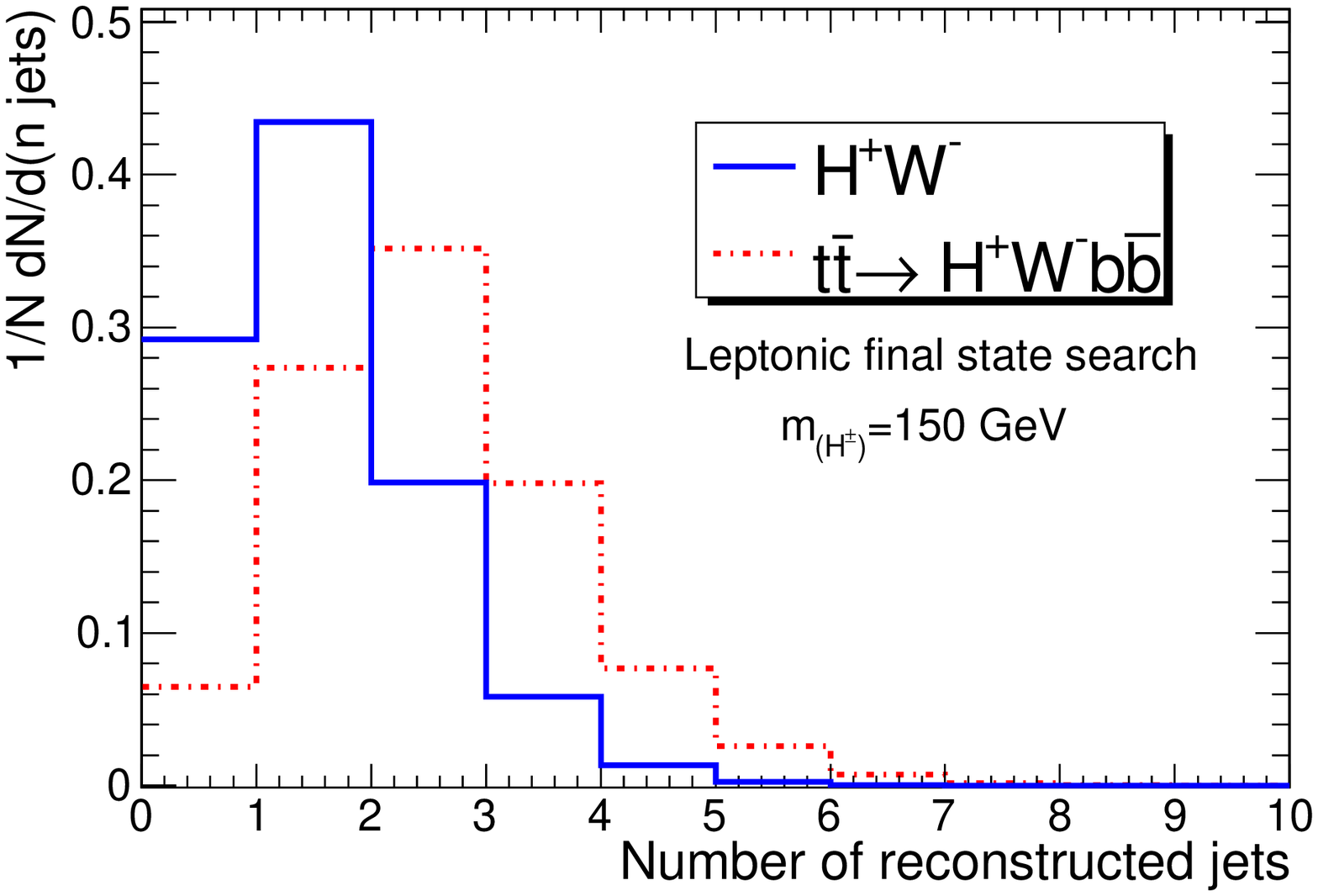}
\end{center}
\caption{Jet Multiplicity in $H^{\pm}W^{\mp}$ and $t\bar{t}\ra H^{\pm}W^{\mp}b\bar{b}$ events in the leptonic final state search with $m_{(H^{\pm})}=150$ GeV.\label{jetmul1}}
\end{figure}

In the following, the contribution of $H^{\pm}W^{\mp}$ to the low mass region search in the two categories of leptonic and hadronic final states is estimated. In other words the additional contribution solely from $H^{\pm}W^{\mp}$ to $t\bar{t}\ra H^{\pm}W^{\mp}b\bar{b}$ which is the main signal in the low mass region is estimated. Having passed a common set of selection cuts, $H^{\pm}W^{\mp}$ signal should appear as an excess of events over what is observed from $t\bar{t}\ra H^{\pm}W^{\mp}b\bar{b}$ events. What is expected is that the final number of $t\bar{t}\ra H^{\pm}W^{\mp}b\bar{b}$ events should be dominant in the signal region due to the large total cross section of this process. One may imagine that a leptonic final state search for events with one jet (the $\tau$ jet) in the final state may produce an $H^{\pm}W^{\mp}$ dominant sample. This is not the case because $t\bar{t}\ra H^{\pm}W^{\mp}b\bar{b}$ events have sizable contribution even to the one-jet bin due to their large cross section at the LHC (Fig. \ref{jetmul1}). To verify this fact the same set of selection cuts as what was used in the heavy charged Higgs leptonic final state search is applied on $H^{\pm}W^{\mp}$ and $t\bar{t}\ra H^{\pm}W^{\mp}b\bar{b}$ events. The chosen parameters for the simulation are $m_{(H^{\pm})}=150$ GeV and $tan\beta=30$.
Table \ref{lowmasstable} lists the selection efficiencies and the remaining number of events after each cut.
\begin{table}
\begin{center}
% use packages: array
\begin{tabular}{|l|c|c|c|c|}
\hline
Process & $H^{\pm}W^{\mp}$ & $t\bar{t}\ra H^{\pm}W^{\mp}b\bar{b}$ \\
\hline
Total cross section [pb] &  0.15 & 878.7     \\
\hline
Number of events at 30 $fb^{-1}$ & 464 & 108111   \\
\hline
N Muons = 1 & 213(45.9$\%$) & 42812(39.6$\%$)   \\
\hline
N Jets = 1 & 93(43.7$\%$) & 11131(26$\%$)   \\
\hline
leading track $p_{T} > 20$ GeV & 50(54$\%$) &  3284(29.5$\%$)  \\
\hline
Isolation &  40(79.2$\%$) & 1438(43.8$\%$) \\
\hline
R $>$ 0.2 & 35(87.7$\%$) & 1214(84.4$\%$)   \\
\hline
1- or 3-prong decay & 34(98.4$\%$) & 1182(97.4$\%$) \\
\hline
Opposite charge & 23(99.9$\%$) & 1158(98$\%$)   \\
\hline
$E^{miss}_{T} > 50$ GeV & 15(66.4$\%$) &  736(63.5$\%$)  \\
\hline
Total efficiency & 4.9$\%$ & 0.67$\%$   \\
\hline
Expected events at $30fb^{-1}$ &  15 & 736   \\
\hline
\end{tabular}
\end{center}
\caption{Selection efficiencies and remaining number of $H^{\pm}W^{\mp}$ and $t\bar{t}\ra H^{\pm}W^{\mp}b\bar{b}$ events after each cut in the $\tau \mu E^{miss}_{T}$ final state. The charged Higgs mass is set to 150 GeV and \tanb = 30. Numbers in parentheses are relative efficiencies in percent with respect to the previous cut. Branching ratios have been taken into account in transition from the second to third row.\label{lowmasstable}}
\end{table}
Therefore as seen from Tab. \ref{lowmasstable}, although the selection efficiency of the $H^{\pm}W^{\mp}$ process is roughly seven times larger they would appear as only $2\%$ excess over $t\bar{t}\ra H^{\pm}W^{\mp}b\bar{b}$ events in the leptonic final state search with \tanb=30.
\subsection{Light Charged Higgs Search, Hadronid Final State}
The contribution of $H^{\pm}W^{\mp}$ to the light charged Higgs search can also be estimated by applying the same selection cuts as was done in the heavy charged Higgs hadronic final state analysis. These cuts are applied on both $H^{\pm}W^{\mp}$ and $t\bar{t}\ra H^{\pm}W^{\mp}b\bar{b}$ events for comparison. Table \ref{lowmasstablehad} shows the selection efficiencies and the remaining number of events after each cut. As seen from Tab. \ref{lowmasstablehad}, $H^{\pm}W^{\mp}$ events appear as a $\sim 1.5\%$ excess over $t\bar{t}\ra H^{\pm}W^{\mp}b\bar{b}$ events in the hadronic final state. Therefore in both categories of leptonic and hadronic searches, in the low mass region, the dominant production process is $t\bar{t}\ra H^{\pm}W^{\mp}b\bar{b}$.
\begin{table}
\begin{center}
% use packages: array
\begin{tabular}{|l|c|c|c|c|}
\hline
Process & $H^{\pm}W^{\mp}$ & $t\bar{t}\ra H^{\pm}W^{\mp}b\bar{b}$ \\
\hline
Total cross section [pb] &  0.15 & 878.7     \\
\hline
Number of events at 30 $fb^{-1}$ & 3020 & 693544  \\
\hline
N Jets = 3 & 800(26.5$\%$) &  2e+05(30.3$\%$)  \\
\hline
leading track $p_{T} > 20$ GeV & 608(76$\%$) & 1.4e+05(67.4$\%$)  \\
\hline
Isolation & 367(60.4$\%$) & 56938(40.2$\%$) \\
\hline
R $>$ 0.2 & 223(60.6$\%$) & 31601(55.5$\%$)   \\
\hline
1- or 3-prong decay & 217(97.4$\%$) & 30463(96.4$\%$) \\
\hline
Exactly one $\tau$ jet & 216(99.7$\%$) & 30402(99.8$\%$)\\
\hline
$W$ mass window & 116(53.7$\%$) & 8664(28.5$\%$) \\ 
\hline
$E^{miss}_{T} > 50$ GeV & 81(69.4$\%$) & 5441(62.8$\%$) \\
\hline
$\Delta\phi_{(\tau,E^{miss}_{T})}$    & 57(70.2$\%$) & 3891(71.5$\%$) \\
\hline
Total efficiency & 1.87$\%$ & 0.56$\%$  \\
\hline
Expected events at $30fb^{-1}$ &  57 & 3891 \\
\hline
\end{tabular}
\end{center}
\caption{Selection efficiencies and remaining number of $H^{\pm}W^{\mp}$ and $t\bar{t}\ra H^{\pm}W^{\mp}b\bar{b}$ events after each cut in the $\tau jj E^{miss}_{T}$ final state search. The charged Higgs mass is set to 150 GeV and \tanb = 30. Numbers in parentheses are relative efficiencies in percent with respect to the previous cut. Branching ratios have been taken into account in transition from the second to third row.\label{lowmasstablehad}}
\end{table}

\section{$5\sigma$ and $3\sigma$ contours}
\begin{figure}
\begin{center}
\includegraphics[width=0.8\textwidth]{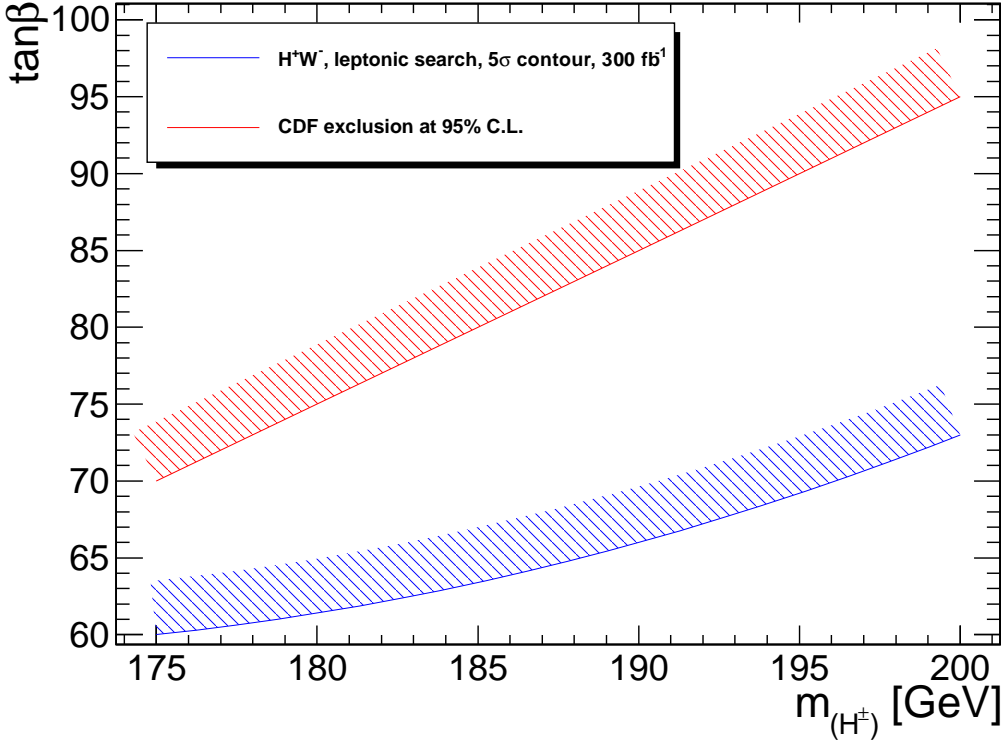}
\end{center}
\caption{The $5\sigma$ discovery contour obtained with the leptonic final state search. The integrated luminosity is set to $300~fb^{-1}$.\label{5sigmalep}}
\end{figure}

\begin{figure}
\begin{center}
\includegraphics[width=0.8\textwidth]{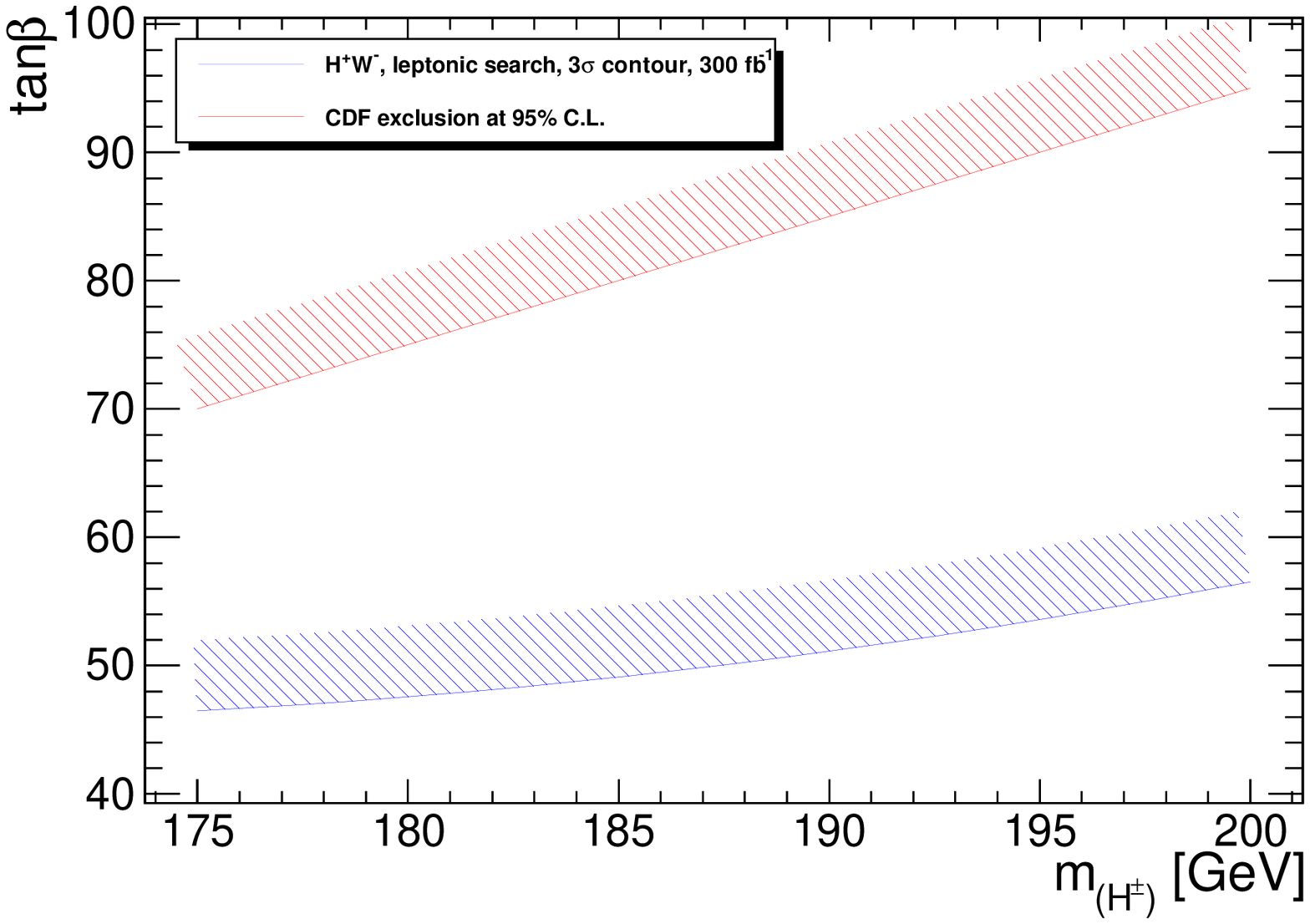}
\end{center}
\caption{The $3\sigma$ evidence contour obtained with the leptonic final state search. The integrated luminosity is set to $300~fb^{-1}$.\label{3sigmalep}}
\end{figure}

\begin{figure}
\begin{center}
\includegraphics[width=0.8\textwidth]{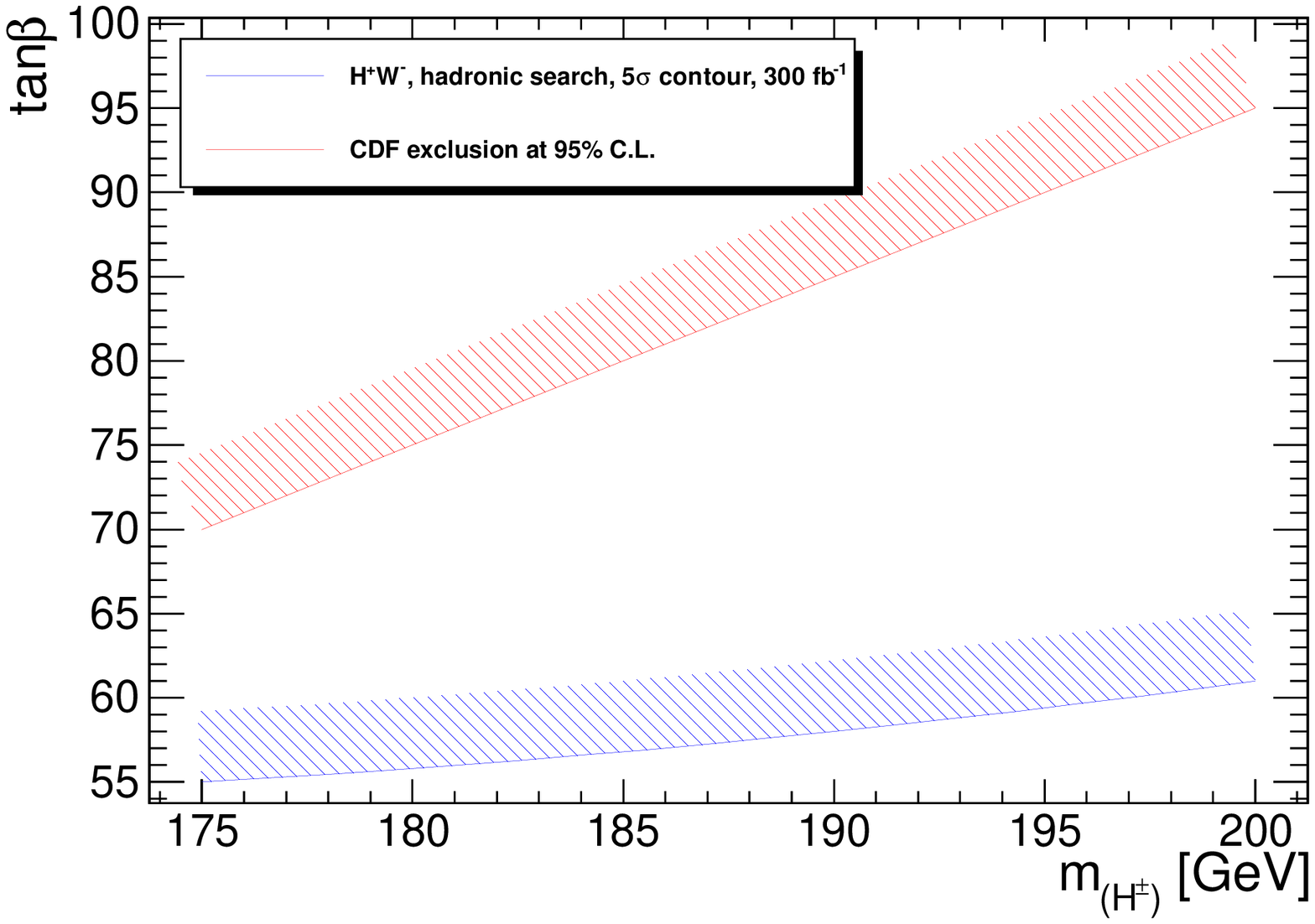}
\end{center}
\caption{The $5\sigma$ discovery contour obtained with the hadronic final state search. The integrated luminosity is set to $300~fb^{-1}$.\label{5sigmahad}}
\end{figure}

\begin{figure}
\begin{center}
\includegraphics[width=0.8\textwidth]{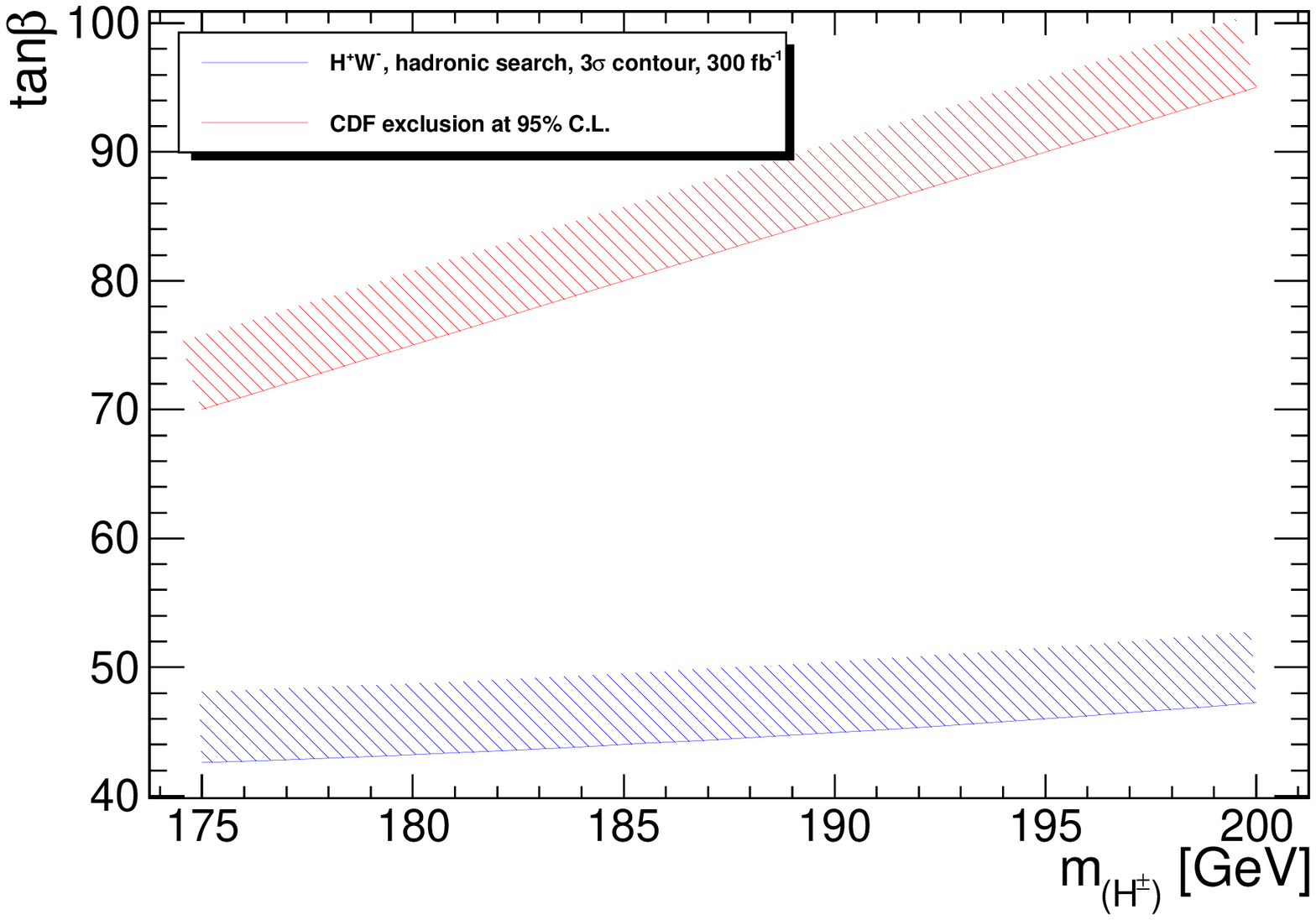}
\end{center}
\caption{The $3\sigma$ evidence contour obtained with the hadronic final state search. The integrated luminosity is set to $300~fb^{-1}$.\label{3sigmahad}}
\end{figure}

In order to obtain $5\sigma$ and $3\sigma$ contours, two more points in the parameter space are studied. They are $m_{(H^{\pm})}=190,~200$ GeV with $tan\beta=30$. Table \ref{2points} shows the cross sections and efficiencies related to these points. It should be noted that the branching ratio of charged Higgs decay to $\tau\nu$ decreases with increasing charged Higgs mass. Using HDECAY, the following values are obtained and used in the analysis: BR($H^{\pm}\ra\tau\nu$)=0.9 for $m_{(H^{\pm})}=190$ GeV and BR($H^{\pm}\ra\tau\nu$)=0.83 for $m_{(H^{\pm})}=200$ GeV. 
\begin{table}
\begin{center}
\begin{tabular}{|c|c||c|c||c|c|}
\hline
$m_{(H^{\pm})}$ & Total cross section & $\textnormal{eff}_{lep.}$ & $N^{lep.}_S$ &  $\textnormal{eff}_{had.}$ & $N^{had.}_S$ \\
\hline
190 GeV & 103 fb & 5.88$\%$ & 35 & 2.74$\%$ & 52 \\
\hline
200 GeV & 95 fb & 6.04$\%$ & 30 & 2.89$\%$ & 46 \\
\hline
\end{tabular}
\end{center}
\caption{Cross sections and efficiencies of leptonic and hadronic final state searches for $m_{(H^{\pm})}=190,~200$ GeV and $tan\beta=30$. $\textnormal{eff}_{lep.}(\textnormal{eff}_{had.})$ is the signal efficiency in leptonic (hadronic) final state search ( the leptonic final state includes both muons and electrons) and $N^{lep.}_S(N^{had.}_S)$ is the number of signal events which pass all the selection cuts at $30~fb^{-1}$.\label{2points}}
\end{table}
It should be noted that although the signal significance stated by Eq. \ref{signif} grows like $\sqrt{L}$ where $L$ is the integrated luminosity, in reality this assumption may not be the case especially when systematic uncertainties are introduced in the signal significance calculation. The estimation of such effects is beyond the scope of this work and needs a full simulation at the presence of detector effects. In this work discovery potential of the studied signal is presented. Figures \ref{5sigmalep} and \ref{3sigmalep} show the estimated $5\sigma$ discovery and $3\sigma$ evidence contours if a leptonic final state search is carried out. Accordingly Figs. \ref{5sigmahad} and \ref{3sigmahad} show the $5\sigma$ discovery and $3\sigma$ evidence contours with the hadronic final state search. As is seen the leptonic and hadronic final state searches turn out to provide almost the same search power for a charged Higgs boson with this channel and the available parameter space is almost the same for both final state searches. The excluded area is the extrapolation of CDF collaboration result in Ref. \cite{cdfexclusion} to the heavy charged Higgs area and high \tanb$~$ values.
\section{Conclusions}
The associated production of charged Higgs and $W$ boson were studied in the framework of MSSM. The signal observability in the leptonic final state search, i.e. $H^{\pm}W^{\mp}\ra \ell~\tau E^{miss}_{T}$ with $\ell=e$ or $\mu$ and also the hadronic final state search, i.e. $H^{\pm}W^{\mp}\ra \tau jj E^{miss}_{T}$ was investigated. The analysis is suitable for a high luminosity study at an integrated luminosity of $300~fb^{-1}$. Results indicate that in the low mass region ($m_{(H^{\pm})}<175$ GeV) this signal only appears as a few percent excess ($2\%$ ($1.5\%$) with $m_{(H^{\pm})}=150$ GeV and \tanb=30 in the leptonic (hadronic) final state) over what is observed from the dominant $t\bar{t}\ra H^{\pm}W^{\mp}b\bar{b}$ process. In the high mass region at an integrated luminosity of $300~fb^{-1}$, a charged Higgs boson with $m_{(H^{\pm})}=175$ GeV, in the leptonic final state search, is in the $5\sigma$ discovery reach if $tan\beta>60$. The $3\sigma$ contour starts from $tan\beta>46$. The hadronic final state search leads to $5\sigma$ signal for $tan\beta>55$ with $m_{(H^{\pm})}=175$. The $3\sigma$ evidence in the hadronic final state search starts from $tan\beta>43$. Although high $tan\beta$ values are accessible with this channel, it can be used to extend the current Tevatron excluded region. Alternatively adding this signal to the current searches at LHC, especially in the high mass search, can provide broader regions in the parameter space than what is currently available and is also able to increase the signal statistics in the transition region. This motivates a full simulation at the presence of detector effects which can be done by CMS and ATLAS collaborations.

\end{document}